\documentclass[twocolumn,aps,prd,superscriptaddress,amsmath,amssymb,nofootinbib,nobibnotes]{revtex4-2}
\usepackage{anyfontsize}
\usepackage{amssymb}
\usepackage{amsmath}
\usepackage{amsfonts}
\usepackage{amsbsy}
\usepackage[utf8]{inputenc} 
\usepackage{newtxtext}
\usepackage{newtxmath}

\usepackage[normalem]{ulem}

\usepackage{tensor}
\usepackage{mathrsfs}
\usepackage{slashed}

\usepackage{bm}

\usepackage[utf8]{inputenc}

\usepackage[usenames,dvipsnames]{xcolor}

\usepackage{verbatim,mathtools,needspace,enumitem,etoolbox}
\usepackage{physics}

\definecolor{linkcolor}{rgb}{0.0,0.3,0.5}

\usepackage[unicode, colorlinks=true, linkcolor=linkcolor, citecolor=linkcolor, filecolor=linkcolor,urlcolor=linkcolor, pdfusetitle]{hyperref}

\usepackage{epstopdf}

\usepackage{bm}
\usepackage{dcolumn}

\usepackage{longtable}
\usepackage{enumerate}
\usepackage[normalem]{ulem}

\hypersetup{colorlinks=true,
citecolor=blue,
linkcolor=blue,
urlcolor=blue}
\usepackage{subfigure}
\usepackage{float}
\usepackage{mathrsfs}
\usepackage[dvipsnames]{xcolor}
\usepackage{soul}

\definecolor{darkgreen}{RGB}{1,212,57}

\usepackage{enumitem}

\usepackage{graphicx}
\usepackage{dcolumn}
\usepackage{bm}

\begin{document}

\title{Twisting shadows: \\ Light rings, lensing and shadows of black holes in swirling universes}

\author{Zeus S. Moreira}
\email{zeus.moreira@icen.ufpa.br}
\affiliation{%
	Programa de Pós-Graduação em Física, Universidade Federal do Pará, 66075-110, Belém, Par{\'a}, Brazil 
}%
\author{Carlos A. R. Herdeiro}
\email{herdeiro@ua.pt}
\affiliation{Departamento de Matem\'atica da Universidade de Aveiro and Centre for Research and Development  in Mathematics and Applications (CIDMA), Campus de Santiago, 3810-183 Aveiro, Portugal.}
\author{Luís C. B. Crispino}
\email{crispino@ufpa.br}
\affiliation{%
	Programa de Pós-Graduação em Física, Universidade Federal do Pará, 66075-110, Belém, Par{\'a}, Brazil 
}%

\begin{abstract}
	Using the Ernst formalism,  a novel solution of \textit{vacuum} general relativity (GR) was recently obtained~\cite{Astorino:2022aam}, describing a Schwarzschild black hole (BH) immersed in a nonasymptotically flat rotating background, dubbed \textit{swirling universe}, with the peculiar property that north and south hemispheres spin in opposite directions. We investigate the null geodesic flow and, in particular, the existence of light rings in this vacuum geometry. By evaluating the total topological charge $w$, we show that there exists one unstable light ring ($w=-1$) for each rotation sense of the background. We observe that the swirling background drives the Schwarzschild BH light rings  \textit{outside} the equatorial plane, displaying counterrotating motion with respect to each other, while (both)  corotating with respect to the swirling universe. Using backwards ray tracing, we obtain the shadow and gravitational lensing effects, revealing a novel feature for observers on the equatorial plane: the BH shadow displays an \textit{odd} $\mathbb{Z}_2$ (north-south) symmetry, inherited from the same type of symmetry of the spacetime itself: a twisted shadow.
\end{abstract}

\maketitle


\section{Introduction}

The first nontrivial exact solution of the vacuum Einstein field equations (EFE), published in 1916 by  Schwarzschild, describes a spherically symmetric, static and asymptotically flat BH spacetime~\cite{Schwarzschild:1916uq}. Nearly five decades later, its rotating generalization was derived by  Kerr~\cite{Kerr:1963ud}. The Kerr solution stands as a cornerstone in BH physics,  hypothesized as the metric describing all BHs in equilibrium - see Ref.~\cite{Herdeiro:2022yle} for a discussion. Moreover, it is also the quintessential solution to learn about rotational effects in GR. Other textbook solutions presenting rotational effects are considered more exotic, such as the Gödel rotating universe~\cite{Godel:1949ga,Hawking:1973uf} and the Taub-NUT spacetime~\cite{Hawking:1973uf,Taub:1950ez,Newman:1963yy}, which contain, for instance, closed timelike curves that are not cloaked by any horizon. By contrast, the Schwarzschild and Kerr metrics are well behaved outside the event horizon, allowing a well defined initial value problem in the outer domain of communication. 

A whole landscape of other exact solutions of the EFE has been derived and cataloged, see e.g. Ref.~\cite{Stephani:2003tm}, some containing rotational effects. But imposing \textit{vacuum} is quite restrictive, making Ricci flat geometries of particular interest. The goal of this paper is to discuss the behavior of light in a class of vacuum solutions of the EFE describing a Schwarzschild BH immersed in a rotating background with the peculiarity that the north and the south hemispheres spin in opposite directions. This \textit{swirling universe} (SU)~\cite{Astorino:2022aam} introduces no closed time-like curves; in this sense it is less exotic than the G\"odel or the Taub-NUT spacetime. Similarly to those, on the other hand, the SU is not asymptotically flat. 

The SU was first mentioned in a work by Gibbons, Mujtaba and Pope~\cite{Gibbons:2013yq}, although its properties were only thoroughly examined in Ref.~\cite{Astorino:2022aam}, wherein the Schwarzschild BH in the SU (SBHSU), as well as its Kerr generalization, were constructed by exploiting the Ernst formalism.\footnote{More recently, new exact solutions involving the SU have been obtained in Refs.~\cite{Barrientos:2024pkt,Astorino:2022prj,Illy:2023iau}.} 
In this paper, we perform an analysis of null geodesics of the SBHSU spacetime. 

Exploring how light bends around compact objects is a crucial aspect of Einstein's theory. The first experimental confirmation of GR was the observation of the gravitational deflection of light during a solar eclipse in 1919~\cite{Dyson:1920cwa,Crispino:2019yew}. The phenomenon of gravitational lensing was investigated by Einstein himself. He studied the case when there is an alignment of the light source, the compact object and the observer, causing the appearance of a ring-like structure, now called Einstein ring~\cite{Einstein:1936llh}. With the landmark observations of the Event Horizon Telescope~\cite{EventHorizonTelescope:2019dse,EventHorizonTelescope:2022wkp}, the study of the gravitational lensing and the observational aspects of ultracompact objects, not only BHs but also BH foils such as boson stars, has become a very active area of research - see e.g. Refs.~\cite{Dabrowski:1998ac,Virbhadra:1999nm,Eiroa:2002mk,Halla:2020hee,Bozza:2010xqn,Cunha:2017wao,Olivares:2018abq,Chowdhuri:2020ipb,Herdeiro:2021lwl,
Afrin:2022ztr,Badia:2022phg,Vagnozzi:2022moj,
Rosa:2022tfv,Chen:2022scf,Rosa:2023qcv,Gao:2023mjb,Wang:2023jop,dePaula:2023ozi,Wen:2022hkv,Atamurotov:2022nim,Hou:2022eev,
Guerrero:2022msp,Kuang:2022xjp,Bezdekova:2022gib,Belhaj:2022cdh,Wang:2021irh}.

Near a BH, light can undergo substantial bending, creating circular trajectories called light rings (LRs). In Refs.~\cite{Cunha:2017qtt,Cunha:2020azh}, two theorems concerning the presence of LRs in generic contexts were established. The first of them demonstrates the existence of pairs of LRs around horizonless ultracompact objects, while the second establishes the existence of LRs specifically for BHs. These theorems do not rely on the Einstein (or other field) equations, but only on appropriate boundary conditions together with a topological technique. Both proofs were formulated based on a set of assumptions that encompass a broad range of spacetimes with physical relevance. An essential assumption is that the spacetime is asymptotically flat. Subsequently, extensions of the LR theorems for different asymptotics were put forward, covering cases such as Schwarzschild de Sitter and anti de Sitter~\cite{Wei:2020rbh}, Schwarzschild-Melvin~\cite{Junior:2021dyw}, Schwarzschild-dilatonic-Melvin~\cite{Junior:2021svb} and Kerr-Newman Taub-NUT~\cite{Wu:2023eml} spacetimes. We note that the SBHSU is not included in the results presented in~\cite{Cunha:2017qtt,Cunha:2020azh}, or any of the other above, due to its peculiar asymptotics. Other related works can be found in Refs.~\cite{Cunha:2024ajc,Ghosh:2021txu}.

Close to a BH, the bending of light combined with its trapping yield a black disk-type image from the perspective of an outside observer, known as the \textit{shadow}~\cite{Falcke:1999pj}, the visual depiction of a BH. The shadow of the Schwarzschild BH observed by a static observer was obtained by Synge~\cite{Synge:1966okc}. The shadow of the Kerr BH was first investigated by Bardeen~\cite{Bardeen:1973tla}. In spacetimes where the geodesic equations cannot be separated, the shadows and gravitational lensing phenomenon can only be analyzed through backward ray tracing methods - see for instance Refs.~\cite{Cunha:2015yba,Cunha:2016bjh,Junior:2021dyw,Junior:2021svb,Sengo:2022jif}.

Here we study the LRs, shadow and gravitational lensing of the SBHSU. As we shall see, the peculiar properties of the SU yield peculiar properties for light propagation, most notably nonequatorial LRs for a single BH spacetime and a BH shadow that  is $\mathbb{Z}_2$-odd, rather than the usual  $\mathbb{Z}_2$ even, i.e. north-south symmetric, when observed from the equatorial plane of the BH spacetime. The remainder of this paper is organized as follows: In Sec.~\ref{Sec. II} we review some general aspects of the Ernst formalism, which is used to obtain the SBHSU solution. We also revisit the main properties of the SBHSU spacetime. In Sec.~\ref{Sec. III} we study the motion of null geodesics of the SBHSU and define the 2-dimensional effective potentials $H_\pm$. We also analyze the LRs structure of the SBHSU, using the techniques developed in Ref.~\cite{Cunha:2020azh}. In Sec.~\ref{Sec. IV} we present our results regarding the shadow and gravitational lensing produced by the SBHSU. Finally, in Sec.~\ref{FR} we present our final remarks. 

\section{The SBHSU spacetime}
\label{Sec. II}
\subsection{The Ernst formalism}
 \label{ErnstFormalism}
The Ernst formalism, developed in the 1960s~\cite{Ernst:1967wx,Ernst:1967by}, has emerged as a powerful mathematical tool for generating stationary and axisymmetric solutions of GR. This methodology proves to be especially valuable when the matter content is solely characterized by the electromagnetic field.\footnote{It is hard to implement the Ernst formalism beyond the electrovacuum case of stationary and axisymmetric spacetimes without spoiling the Ernst equation symmetries. Nevertheless, there are a few possible extensions, such as also considering a minimally and conformally coupled scalar field~\cite{Astorino:2013xc}.} The central idea of this formalism lies in rewriting the Einstein-Maxwell (EM) system of equations in terms of the Ernst potentials. The resulting equations, referred to as the Ernst equations, allow for a clearer exploration of the theory's symmetries, thereby facilitating a structured examination of group theoretical aspects of the EM system. 

For the stationary vacuum case, Geroch showed that $SU(1,1)$ is the underlying group of symmetries of the Ernst equations~\cite{Geroch:1970nt}. Subsequently, Kinnersley demonstrated that, when generalized for electrovacuum, the corresponding symmetry group linked to the Ernst equations is isomorphic to $SU(2,1)$~\cite{10.1063/1.1666373,Kinnersley:1977pg}. Furthermore, this symmetry group can be explicitly represented through five general transformations. Starting from one known  solution, it is possible, through Kinnersley transformations, to generate nontrivial new solutions of the EM theory.

Considering the Schwarzschild BH as a  starting point (seed), using the Kinnersley transformations and the conjugation discrete transformation (see Eq.~\eqref{con} below), it is possible to generate several other solutions like:  Reissner-Nordström, Taub-NUT and the Schwarzschild-Melvin BHs. Following this approach, Astorino, Martelli, and Viganò showed that one can obtain the Schwarzschild and Kerr BHs immersed in a SU \cite{Astorino:2022aam}. They obtained these BHs solutions by means of the Ehlers transformations. The investigation conducted in Ref. \cite{Astorino:2022aam} goes beyond the metric derivation and includes also analyses of the horizon embedding, ergoregion, closed timelike curves, conical singularities and geodesics. 

Applying the Ernst formalism requires both stationarity and axisymmetry. Stationarity means that there must exist a  (asymptotic) timelike Killing vector field $\xi$; axisymmetry implies another Killing vector field $\psi$, whose trajectories form closed spacelike curves. We also assume that the two Killing vector fields commute,\footnote{We remark that in asymptotically flat spacetimes, there is a theorem established by Carter \cite{Carter:1970ea} which guarantees the commutation of these two Killing vector fields for asymptotically flat spacetimes. However, we will not consider asymptotically flatness as a hypothesis here.} which implies that we have the freedom to select coordinates $(t, \varphi)$ that are specifically suited to these symmetries. In this coordinate system, $\xi$ corresponds to $\partial_t$, and $\psi$ corresponds to $\partial_\varphi$, as established in Ref.~\cite{Hawking:1973uf}. Additionally, the metric components remain independent of these chosen coordinates.

The spacetime to be constructed is a solution of the EM equations, without the inclusion of a cosmological constant. This system of equations is expressed as follows:

\begin{equation}\label{EM1}
	R_{\mu\nu} - \frac{1}{2}Rg_{\mu\nu} = 2\left(F_{\mu\alpha}F_\nu{}^\alpha - \frac{1}{4}g_{\mu\nu}F_{\alpha\beta}F^{\alpha\beta}\right),
\end{equation}

\begin{equation}\label{EM2}
	\partial_\mu\left(\sqrt{-g}F^{\mu\nu}\right) = 0,
\end{equation}
where $R_{\mu\nu}$ is the Ricci tensor and $R$ the Ricci scalar constructed from the metric $g_{\mu\nu}$. The Maxwell-Faraday tensor is represented by $F_{\mu\nu}$ and is defined by the $U(1)$ gauge potential $A_\mu$ according to $F_{\mu\nu}=\partial_\mu A_\nu-\partial_\nu A_\mu$. Assuming that the gauge potential inherits the symmetries from the spacetime, we must have $A=A_tdt+A_\varphi d\varphi$.

Within this framework, the most general line element is described bt the Lewis-Weyl-Papapetrou (LWP) metric, given by
\begin{equation}\label{lwp}
	ds^2=-f \left(dt-\omega d\varphi\right)^2+f^{-1}\left[\rho^2d\varphi^2+e^{2\gamma}\left(d\rho^2+dz^2\right)\right].
\end{equation}
In Eq.~\eqref{lwp}, we have adopted the Weyl canonical coordinates $(t, \rho, z, \varphi)$, where $z\in(-\infty,\infty)$ and $\rho\in[0,\infty)$. The function $\omega$ is associated to the spacetime rotation with respect to the axis $\rho=0$. By substituting the LWP metric back into the EM system, one may obtain a set of four coupled, partial differential equations for the functions $f$, $\omega$, $A_t$ and $A_\varphi$ (see Refs.~\cite{Etevaldo,Vigano:2022hrg} for more details). The equations for $\gamma$ decouple from the others, implying that $\gamma$ is fully determined by the remaining functions.

Let $\mathcal{E}$ and $\Phi$ be complex functions representing the Ernst potentials defined according to
\begin{equation}
	\Phi=A_t+i\bar{A}_\varphi,
\end{equation}
\begin{equation}
	\mathcal{E}=f-\left|\Phi\Phi^*\right|+ih,
\end{equation}
where 
\begin{equation}\label{a}
	\nabla \bar{A}_\varphi=f\rho^{-1}\hat{e}_\varphi\times(\nabla A_\varphi+\omega \nabla A_t),
\end{equation}
\begin{equation}\label{h}
	\nabla h=-f^2\rho^{-1}\hat{e}_{\varphi}\times\nabla\omega-2\text{Im}(\Phi^*\nabla\Phi).
\end{equation}
The operator $\nabla$ in Eqs.~\eqref{a} and~\eqref{h} is the flat vectorial operator associated with the ``nonphysical'' metric $ds^2=d\rho^2+dz^2+\rho^2d\varphi^2$, written in cylindrical Weyl coordinates. The vector $\hat{e}_\varphi$ corresponds to the unit vector in the azimuthal direction. 

Ernst showed that Eqs.~\eqref{EM1} and~\eqref{EM2}, assuming the previously mentioned spacetime symmetries, are equivalent to the Ernst equations, which are given by
\begin{equation}\label{EE1}
	\left(\Re\mathcal{E}+\left|\Phi\right|^2\right)\nabla^2\mathcal{E}=\left(\nabla\mathcal{E}+2\Phi^*\nabla\Phi\right)\cdot\nabla\mathcal{E},
\end{equation}
\begin{equation}\label{EE2}
	\left(\Re\mathcal{E}+\left|\Phi\right|^2\right)\nabla^2\Phi=\left(\nabla\mathcal{E}+2\Phi^*\nabla\Phi\right)\cdot\nabla\Phi.
\end{equation}

The Ernst equations are symmetric under the action of the group $SU(2,1)$, which, following Kinnersley, can be represented by five transformations on the Ernst potentials. The Kinnersley transformations are given by~\cite{10.1063/1.1666373,Kinnersley:1977pg}

\begin{widetext}
	\begin{align} 
		\label{K1} 
		\mathcal{E}\rightarrow\mathcal{E}'= \alpha \alpha^* \mathcal{E}\, ,\ \ \  &\Phi\rightarrow\Phi'= \alpha \Phi\, , \\ 
		\label{K2} \mathcal{E}\rightarrow\mathcal{E}'=\mathcal{E}+ib\, ,\ \ \  & \Phi\rightarrow\Phi'=\Phi\, , \\ 
		\label{K3} \mathcal{E}\rightarrow\mathcal{E}'=\frac{\mathcal{E}}{1+ic\mathcal{E}}\, ,\ \ \  &\Phi\rightarrow\Phi'=\frac{\Phi}{1+ic\mathcal{E}}\, ,\\ 
		\label{K4} \mathcal{E}\rightarrow\mathcal{E}'=\mathcal{E}-2\beta\Phi-\beta\beta^*\, ,\ \ \  &\Phi\rightarrow\Phi'=\Phi+\beta^*\, ,\\ 
		\label{K5}
		\mathcal{E}\rightarrow\mathcal{E}'=\frac{\mathcal{E}}{1-2\gamma^*\Phi-\gamma\gamma^*\mathcal{E}}\, ,\ \ \  &\Phi\rightarrow\Phi'=\frac{\Phi+\gamma\mathcal{E}}{1-2\gamma^*\Phi-\gamma\gamma^*\mathcal{E}}\, ,
	\end{align}
\end{widetext}
where $b, c \in \mathbb{R}$ and $\alpha,\beta,\gamma\in \mathbb{C}$. Hence, the aforementioned transformations depend upon 8 arbitrary real parameters, which matches the dimension of $SU(2,1)$\footnote{Let $GL(3,\mathbb{C})$ be the general linear group of order $3$ over $\mathbb{C}$ and $\eta=\text{diag}(1,1,-1)$. Since $SU(2,1)\{A\in GL(3,\mathbb{C})|\det A=1\ \text{and } A\eta A^\dagger=\eta\}$, the dimension of $SU(2,1)$ must be 8, since we have 18 real free parameters from  $GL(3,\mathbb{C})$ that are constrained by 1+9 conditions according to $\det A=1$ and $A\eta A^\dagger=\eta$, respectively.}. Notably, these transformations map the space of solutions of the EM equations into itself. We remark, however, that Eqs.~\eqref{K1},~\eqref{K2} and~\eqref{K4} lead to trivial transformations, i.e. the modifications can be absorbed in gauge transformations of either the metric or the electromagnetic potential. The transformations defined in Eqs.~\eqref{K3} and~\eqref{K5} are known as Ehlers and Harrison transformations, respectively, and act on the Ernst potential in a nontrivial way, giving rise to nonequivalent spacetimes solutions.
\subsection{Conjugate metrics}
There is a property of the LWP metric that is, if $(f,\omega,\gamma)$ constitutes a triple defining a LWP line element that solves the EM system, then there exist $(\bar{f},\bar{\omega},\bar{\gamma})$ defining a nonequivalent LWP metric which is also a solution. These triples must be related by a discrete transformation called conjugation, defined by~\cite{Chandrasekhar:1985kt,Astorino:2022aam}
\begin{equation}\label{con}
	\left\{f\rightarrow \frac{\rho^2}{\bar{f}}-\bar{f}\bar{\omega}^2,\omega\rightarrow\frac{\bar{f}^2\bar{\omega}}{\bar{f}^2\bar{\omega}^2-\rho^2},e^{2\gamma}\rightarrow e^{2\bar{\gamma}}\left(\frac{\rho^2}{\bar{f}^2}-\bar{\omega}^2\right)\right\}.
\end{equation}

After applying the transformation given in Eq.~\eqref{con}, one may obtain the metric
\begin{equation}\label{lwp2}
	d\bar{s}^2=-\bar{f}\left(d\varphi-\bar{\omega} dt\right)^2+\bar{f}^{-1}\left[\rho^2dt^2+e^{2\bar{\gamma}}\left(d\rho^2+dz^2\right)\right].
\end{equation}

The two LWP metrics correspondent to Eqs.~\eqref{lwp} and~\eqref{lwp2}, related by Eq.~\eqref{con}, are called conjugate metrics. 

\begin{figure}[h!]
	\centering
	\includegraphics[width=\columnwidth]{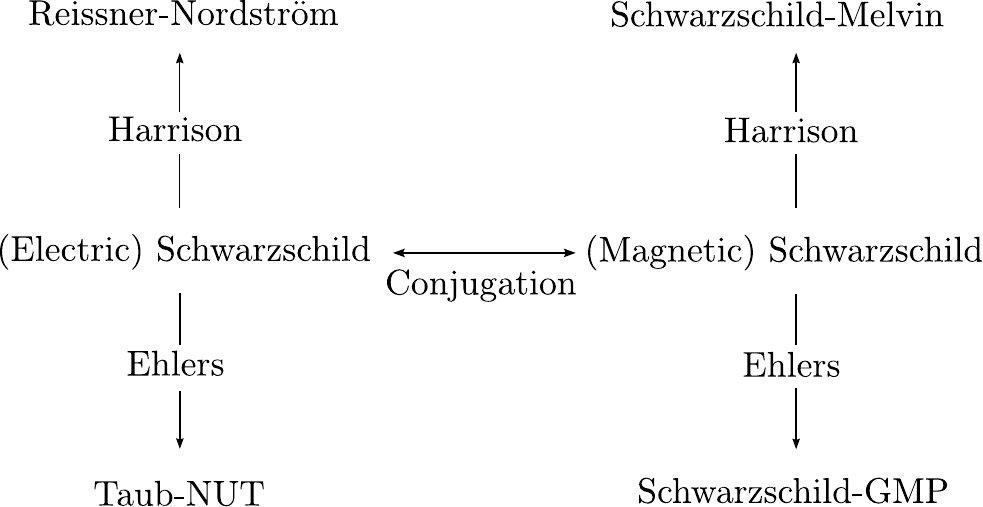}
	\caption{Schematic representation of the metrics generated from Schwarzschild spacetime using conjugation and Kinnersley transformations.}
	\label{transformations}
\end{figure}

The combination of Kinnersley transformations obtained from the Ernst formalism, together with the conjugation operation, enables one to obtain a total of five nonequivalent metrics from only one seed. In Ref.~\cite{Astorino:2022aam} the authors refer to the metric of Eq.~\eqref{lwp} as electric LWP metric, whereas the metric of Eq.~\eqref{lwp2} is referred to as magnetic LWP metric. In Fig.~\ref{transformations} we display all solutions that can be generated starting with Schwarzschild spacetime as a seed.

From the electric version of Schwarzschild metric we may obtain Reissner-Nordström (Taub-NUT) by means of a Harrison (Ehlers) transformation. If, instead, we first conjugate the Schwarzschild solution to obtain the magnetic version of the Schwarzschild BH, it is possible to generate the Schwarzschild-Melvin (SBHSU) metric through a Harrison (Ehlers) transformation.

\subsection{The metric}\label{SecMet}
As discussed in the previous subsection, the SBHSU spacetime can be obtained by means of an Ehlers transformation on the magnetic Schwarzschild solution. It is an algebraically general, stationary, axially symmetric and nonasymptotically flat BH solution of the vacuum Einstein equations. Its line element can be written as 
\begin{equation}
	\begin{aligned}\label{BHmetric}
		ds^2=F(r,\theta)\bigg[-f(r)dt^2&+\frac{dr^2}{f(r)}+r^2d\theta^2\bigg]+\frac{r^2\sin^2\theta}{F(r,\theta)}\times\\
		&\Big\{d\varphi+\left[4jrf(r)\cos\theta\right]dt\Big\}^2,
	\end{aligned}
\end{equation}
with 
\begin{equation}
	f(r)=1-\frac{2M}{r},
\end{equation}
\begin{equation}
	F(r,\theta)=1+j^2r^4\sin^4\theta.
\end{equation}

For stationary, axisymmetric and asymptotically flat BH spacetimes, the central object mass and angular momentum may be defined via Komar integrals~\cite{Komar:1963svp}. These integrals are given by

\begin{equation}
	M_{\text{Komar}}=-\frac{1}{8\pi}\lim\limits_{r\rightarrow \infty}\int_{\mathbb{S}^2}\star d\xi^\flat,
\end{equation}
\begin{equation}
	J_{\text{Komar}}=\frac{1}{16\pi}\lim\limits_{r\rightarrow \infty}\int_{\mathbb{S}^2}\star d\psi^\flat,
\end{equation}
where $\mathbb{S}^2$ represents a topological spherical 2-surface, $\star$ is the Hodge star operator and $\flat$ is the musical isomorphism from the tangent to the cotangent bundle \cite{lee2003introduction}.

For nonasymptotically flat spacetimes, even if the Komar integrals can be defined, their physical interpretation is no longer straightforward. For the SBHSU, on $\mathbb{S}^2$, we have

\begin{equation}
	\begin{aligned}
		\star d\xi^\flat=r^2 \sin\theta&\bigg[\frac{4j^2r^3f(r)\sin^22\theta}{F(r,\theta)^2}\left(f(r)+rf'(r)\right)\\&-f(r)\frac{\partial_r F(r,\theta)}{F(r,\theta)}-f'(r)\bigg]d\theta\wedge d\varphi,
	\end{aligned}
\end{equation}
\begin{equation}
	\star d\psi^\flat=\frac{4 \mathit{j} r^4 \sin ^3\theta \cos\theta \left(r f'(r)+f(r)\right)}{F(r,\theta )^2}d\theta\wedge d\varphi,
\end{equation}
which, after the integration over $\mathbb{S}^2$, results in
\begin{equation}
	M_{\text{Komar}}=M, \ \ \ \ J_{\text{Komar}}=0.
\end{equation}

This calculation shows that $M$ can be heuristically interpreted as the BH mass and the total angular momentum is zero. Although $J_{\text{Komar}}=0$, the space is not static (see Sec.~\ref{ErgoSec}). The Authors in Ref.~\cite{Astorino:2022aam} obtained the same result using the canonical integrability method.

When $j=0$, the SBHSU metric simplifies to the Schwarzschild solution, and when $M=0$, it reduces to the SU, whose line element can be expressed as
\begin{equation}\label{Smetric}
	\begin{aligned}
		ds^2=&(1+j^2\rho^4)(-dt^2+d\rho^2+dz^2)\\&+\frac{\rho^2}{1+j^2\rho^4}(d\varphi+4jzdt)^2,
	\end{aligned}
\end{equation}
where we have written Eq.~\eqref{Smetric} in cylindrical coordinates
\begin{equation}\label{cylindrical}
	\rho=r\sin\theta,\ \ \ \ z=r\cos\theta.
\end{equation}
The parameter $j$ is associated with the background spacetime rotation.

The SBHSU is not plagued with geometric pathologies such as conical singularities, nor causality violation due to closed timelike curves~\cite{Astorino:2022aam}. Nevertheless, the SBHSU solution has a singularity at $r=2M$ and $r=0$, inherited from the Schwarzschild BH. The former  is a coordinate singularity and defines the location of the event horizon, whereas the latter cannot be removed by a change of coordinates, since the curvature invariant, given by
\begin{equation}\label{curvature}
	R_{\mu\nu\alpha\beta}R^{\mu\nu\alpha\beta}=\frac{\mathcal{F}(r,\theta)}{r^6},
\end{equation} 
diverges as $r\rightarrow 0$, for all $\theta\in(0,\pi)$, indicating that the spacetime is, indeed, singular. The function $\mathcal{F}(r,\theta)$ is a lengthy expression in terms of the coordinates $r$ and $\theta$ (which we choose not to show explicitly in this work) that satisfies $\lim\limits_{r\rightarrow 0}\mathcal{F}(r,\theta)=48M^2$. 

On the other hand, the background spacetime defined by Eq.~\eqref{Smetric} is everywhere regular, as it can be shown by setting $M=0$ in Eq.~\eqref{curvature}, obtaining
\begin{equation}
	R_{\mu\nu\alpha\beta}R^{\mu\nu\alpha\beta}|_{M=0}=\frac{192 j^2 \left(j^6 \rho ^{12}-15 j^4 \rho^8+15 j^2 \rho ^4-1\right)}{\left(j^2 \rho^4+1\right)^6}.
\end{equation}

\begin{figure}[h!]
	\centering
	\includegraphics[width=\columnwidth]{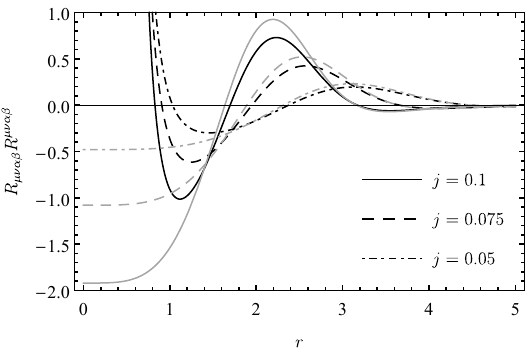}
	\caption{Kretschmann scalar of the SU (gray) and SBHSU (black) spacetimes, for $\theta=\pi/2$ and $j=0.05,\ 0.075,\ 0.1$. 
	For the BH cases we have set $M=0.1$ and $\theta=\pi/2$.}
	\label{KretschmannPlot}
\end{figure}

In Fig.~\ref{KretschmannPlot} we plot the Kretschmann scalar, setting $\theta=\pi/2$, for SU and SBHSU spacetimes. In both cases, the curvature scalar tends to zero as $r\to\infty$; however, while $|R_{\mu\nu\alpha\beta}R^{\mu\nu\alpha\beta}|<\infty$ as $r\to 0$ for the SU, the Kretschmann scalar diverges for the SBHSU.

\subsection{Ergoregion}\label{ErgoSec}
\begin{figure}
	\centering
	\includegraphics[width=\columnwidth]{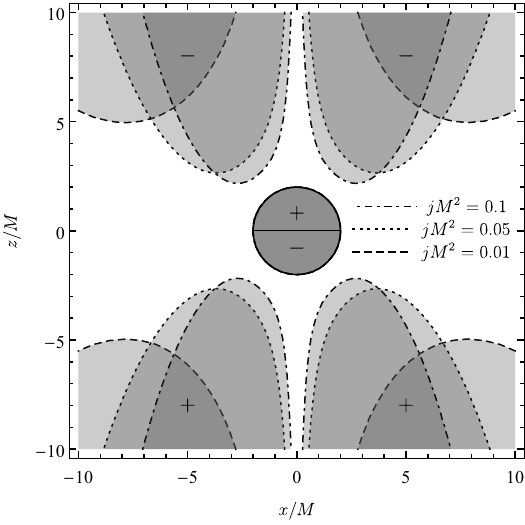}
	\caption{Representation of the surface $g_{tt}(r,\theta)=0$ in the section $y=0$ of the 3-dimensional Euclidean space, for $jM^2=0.01,0.05,0.1$. As the parameter of the swirling background increases, the ergosurface gets closer to the horizon surface.}
	\label{ErgoRegion}
\end{figure}

This solution also possesses an ergoregion, implicitly defined by $g_{tt}(r,\theta)=0$ (excluding the surface $r=2M$)~\cite{Astorino:2022aam,Capobianco:2023kse}. We consider the $g_{tt}$ metric component as a function $g_{tt}:\mathbb{R}^3\rightarrow\mathbb{R}$, where $r$ and $\theta$ are the usual spherical polar coordinates in $\mathbb{R}^3$. Let $\{x,y,z\}$ be the rectangular coordinates in $\mathbb{R}^3$. We plot in Fig.~\ref{ErgoRegion} the cross-section through the surface $y=0$ of the surface $g_{tt}(r,\theta)=0$ (which is not an isometric embedding).

The ergosurface is defined by 2 disconnected patches, namely: the two noncompact branches above and below the BH. The direction of the frame-dragging can be inferred from

\begin{equation}\label{drag}
	\frac{d\phi}{dt}=-\frac{g_{t\varphi}}{g_{\varphi\varphi}}=-4j(r-2M)\cos\theta \, .
\end{equation}

Therefore, inside the top (bottom) branch, as well as in the region inside the BH, below (above) the equatorial plane, observers must rotate in the negative (positive) direction. The sense of rotation is  also indicated in Fig.~\ref{ErgoRegion} by $+$ ($-$) sign if it is positively (negatively) oriented. The opposite spin directions of the north and south is what renders the zero angular momentum obtained in Sec.~\ref{SecMet}. Similar to the SBHSU case, the Reissner-Nördstrom-Melvin solution also has a noncompact ergoregion~\cite{Gibbons:2013yq}.

We remark that the Killing vector $\xi=\partial_t$ is asymptotically timelike in the sense that $g_{tt}(r,\theta)\underset{r\to\infty}{\approx}-j^2r^4\sin^4\theta$. This quantity is negative for every $\theta\in(0,\pi)$, but tends to zero when $\theta\to 0$ or $\theta\to \pi$. Hence, in the vicinity of the symmetry axis, $\xi$ may lose its timelike character~\cite{Astorino:2022aam}. This serves as an indicator of the presence of a noncompact ergoregion in the vicinity of the rotational axis.

\subsection{Horizon geometry}
The geometry of the event horizon is defined by Eq.~\eqref{BHmetric} restricted to the 2-surface $t=$constant and $r=2M$, from where we obtain
\begin{equation}\label{Hmetric}
	ds^2|_{\substack{r=2M \\ t=\text{const}}}= 4M^2F(2M,\theta)d\theta^2 +\frac{4M^2 \sin^2\theta}{F(2M,\theta)}d\varphi^2.
\end{equation} 

For a two-dimensional surface embedded in a three-dimensional space, the Gaussian curvature $K$ is defined as the product of the principal curvatures $k_1$ and $k_2$ at each point. The principal curvatures represent the maximum and minimum curvatures in orthogonal directions on the surface. In our case, the two principal directions are $\{\theta,\varphi\}$ and we will represent the corresponding principal curvatures by $\{k_\theta,k_\varphi\}$, respectively. From Eq.~\eqref{Hmetric} we can compute the Gaussian curvature~\cite{Manfredo}
\begin{equation}
	\begin{aligned}
		K&=-\frac{1}{2\sqrt{\det (g_{\mu\nu})}}\frac{\partial}{\partial\theta}\left(\frac{\partial_\theta g_{\varphi\varphi}}{\sqrt{\det (g_{\mu\nu})}}\right)\\
		&=\frac{\mathcal{K}(\theta)}{4M^2(1+16j^2M^4\sin^4\theta)^3},
	\end{aligned}
\end{equation}
where $g_{\mu\nu}$ are the metric components of Eq.~\eqref{Hmetric} and
\begin{equation}
\begin{aligned}
		\mathcal{K}(\theta)=1+48 j^2 M^4 \sin ^22 \theta -1024 &j^4 M^8 \sin ^6\theta  \cos ^2\theta\\
		&-256 j^4 M^8 \sin ^8\theta.
\end{aligned}
\end{equation}

One key significance of the Gaussian curvature lies in its classification of different types of surfaces regarding its sign. By examining the signs and values of the Gaussian curvature at various points on a surface, we can identify whether the surface is flat/parabolic\footnote{In both flat and parabolic points, the Gaussian curvature vanishes, but in flat spaces both principal curvatures are zero, whereas in parabolic points just one of the principal curvatures is zero. An example of a surface with parabolic points is the surface of a cylinder.}, positively or negatively curved. 
One can show that, for all $j>0$, $\theta=\pi/2$ corresponds to a global minimum of $K$, where it is evaluated to
\begin{equation}
	K|_{\theta=\pi/2}=\frac{1-16 j^2 M^4}{4M^2 \left(1+16 j^2 M^4\right)^2}.
\end{equation}

Therefore, for $j<1/4M^2$ the Gaussian curvature is always positive. If we set $j=1/4M^2$, the Gaussian curvature is zero at the equatorial plane. If $j$ is greater than this critical value, then $K$ is negative on a neighborhood of $\theta=\pi/2$. Since $k_\varphi$ is always positive due to axisymmetry, $k_\theta$ must be negative in a neighborhood of $\theta=\pi/2$ for $j>1/4M^2$. 

We can classify the horizon geometry according to the sign of the Gaussian curvature in the following way~\cite{Manfredo}:
\begin{itemize}
	\item[(i)] $j<1/4M^2$: Globally elliptic;
	\item[(ii)] $j=1/4M^2$: Everywhere elliptic, except at the equator, where it is parabolic;
	\item[(iii)] $j>1/4M^2$: Locally elliptic at the poles and locally hyperbolic at the equator.
\end{itemize}

\begin{figure}
	\centering
	\includegraphics[width=\columnwidth]{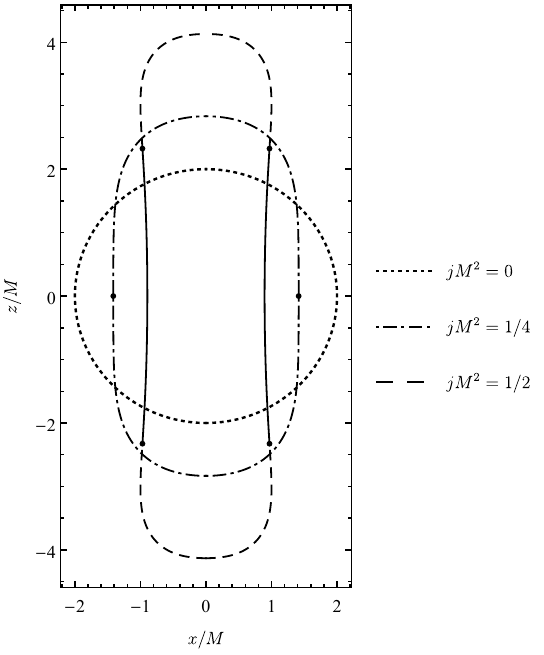}
	\caption{Isometric embedding of the horizon into the 3-dimensional Euclidean space, for $jM^2=0,1/4,1/2$. As the parameter of the swirling background increases, the horizon gets more prolate and thinner along the equator.}
	\label{Embedding}
\end{figure}

In Fig.~\ref{Embedding} we plot the isometric embedding for $jM^2=0,1/4,1/2$, specifying where $K$ is zero (negative) by points (continuous lines). The dashed, dot dashed and dotted lines represent points where $K$ is greater than zero.

Another way to deduce the horizon local hyperbolic geometry at equator for $j>1/4M^2$ is to calculate the proper length 
\begin{equation}
	\begin{aligned}
		L(\theta)&=\int_{0}^{2\pi}\sqrt{g_{\varphi\varphi}(r,\theta)}|_{r \to 2M}d\varphi\\
		&=2\pi \sqrt{\frac{4  M^2 \sin ^2\theta }{1+16 j^2 M^4 \sin ^4\theta}} \ ,
	\end{aligned}
\end{equation}
which represents the horizon perimetral length at a fixed value of $\theta$.

For every compact, globally elliptic and $\mathbb{Z}_2$ symmetric surface contained in $\mathbb{R}^3$, the length $L(\theta)$  attains its maximum at $\theta=\pi/2$. This holds true for $j<1/4M^2$. However, as we extend to $j>1/4M^2$, the point $\theta=\pi/2$ transitions to a local minimum of $L(\theta)$, while the points $\arcsin\sqrt{1/4jM^2}$ and $\pi-\arcsin\sqrt{1/4jM^2}$ become maxima. 

The existence of an elliptic portion on the horizon geometry is expected, since every compact surface embedded in $\mathbb{R}^3$ has, at least, one elliptical point~\cite{Manfredo}. Thus, either the horizon surface is purely elliptical or there exists a region where it can be hyperbolic. In the Kerr case, for high enough values of the dimensionless rotation parameter ($a/M>\sqrt{3}/2$), there are regions where the horizon surface is hyperbolic,
namely, in a neighborhood of the poles. In the vicinity of those points, the isometric embedding into the Euclidean 3-space is not realizable. Generically, any $U(1)$-symmetric $2d$-surface with negative Gaussian curvature at  fixed points of a $U(1)$ symmetry cannot be globally embedded into Euclidean 3-space~\cite{Frolov:2006yb}. In the case of SBHSU, the negative curvature regions are not vicinities of the poles, hence, the previous theorem does not apply. In fact, the global embedding exists for all values of $j$.

Although the horizon gets deformed with increasing $j$, the horizon area $\mathcal{A}$ does not change, i.e. the area of the Schwarzschild BH $\mathcal{A}=16\pi M^2$ is preserved, regardless of $j$.

\section{LRs in the SBHSU (and SU) spacetimes}
\label{Sec. III}
\subsection{Null geodesics}
We now  study null geodesics on the SBHSU. We adopt the Hamiltonian formalism. In order to do so, we consider the Hamiltonian function $\mathscr{H}$, 
\begin{equation}
	\mathscr{H}=\frac{1}{2}g^{\mu\nu}p_{\mu}p_{\nu}=0,
\end{equation}
where $p_\mu$ is the 4-momentum of the photon. The corresponding Hamilton's equations are given by
\begin{equation}\label{xdot}
	\dot{x}^\mu=\frac{\partial\mathscr{H}}{\partial p_\mu},
\end{equation}
\begin{equation}\label{pdot}
	\dot{p}_\mu=-\frac{\partial\mathscr{H}}{\partial x^\mu}.
\end{equation}
In Eqs.~\eqref{xdot} and~\eqref{pdot} the dots stands for derivatives with respect to an affine parameter.

As discussed in Sec.~\ref{ErnstFormalism}, the SBHSU admits two Killing vectors fields, namely, $\partial_t$ and $\partial_\varphi$. These vectors are responsible for generating the isometries governing time translation and rotation around the symmetry axis, respectively. The existence of the two Killing vectors also implies the existence of two quantities, defined by
\begin{equation}
	E=-(\partial_t)^\mu p_\mu=-p_t,
\end{equation}
\begin{equation}
	L=(\partial_\varphi)^\mu p_\mu=p_\varphi,
\end{equation}
which are conserved along the direction of the geodesics.

The Hamiltonian can be separated into a kinetic term $T$ plus a potential $V$, according to
\begin{equation}
	\mathscr{H}=T+V,
\end{equation}
where
\begin{equation}
	T=g^{rr}(p_r)^2+g^{\theta\theta}(p_\theta)^2,
\end{equation}
\begin{equation}
	V=	\frac{L^2  F(r,\theta )}{r^2\sin^2\theta}-\frac{\left(4 j r f(r)L \cos \theta +E\right)^2}{f(r) F(r,\theta )}.
\end{equation}

It is possible to define potentials $H_{\pm}$, which do not depend on the conserved quantities $E$ and $L$. Such potentials are given by

\begin{figure*}
	\centering
	\includegraphics[scale=1]{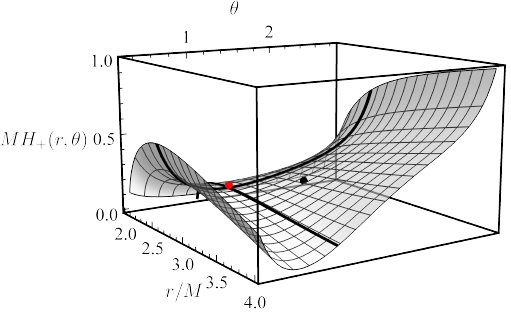}
	\includegraphics[scale=1]{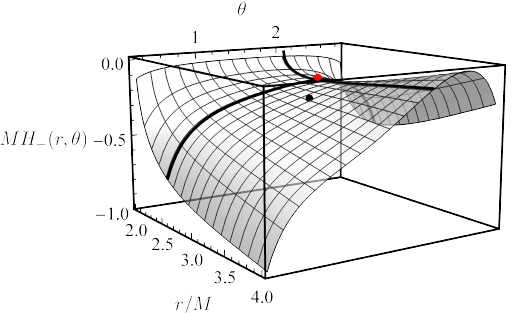}
	\caption{3D plot representations showing the potentials $H_{+}$ (left panel) and $H_{-}$ (right panel), as a function of the $(r,\theta)$ coordinates for $M^2 j=0.05$. The critical points for both $H_\pm$ are highlighted in red, while the Schwarzschild LR locations are denoted by black points.}
	\label{Potential}
\end{figure*}

\begin{equation}\label{Hpm}
	H_{\pm}(r,\theta)=\pm \frac{\sqrt{f(r)} F(r,\theta)}{r\sin\theta}-4 j r f(r)\cos\theta,
\end{equation}
and are related with $V$ by
\begin{equation}\label{V}
	V=-\frac{L^2}{f(r)F(r,\theta)}\left(\frac{E}{L}-H_{+}\right)\left(\frac{E}{L}-H_{-}\right).
\end{equation}
In Fig.~\ref{Potential} we illustrate the null geodesic potentials $H_{\pm}$.

We remark that the potentials $H_{\pm}(r,\theta)$, defined in Eq.~\eqref{Hpm} satisfy 

\begin{equation}\label{Hsym}
	H_{\pm}(r,\theta)=-H_{\mp}(r,\pi-\theta),
\end{equation}
which implies that the potential $H_{-}$ can be fully constructed from $H_+$ and vice-versa.

\subsection{Topological charge}
We now inspect the LRs of the SBHSU. By definition, LRs are critical points of the potential $V$ defined in Eq.~\eqref{V}. In Ref.~\cite{Cunha:2017qtt} it was shown that critical points of $V$ are also critical points of $H_\pm$. The critical points $(r_\pm,\theta_\pm)$ of $H_{\pm}$ for the SBHSU with $jM^2=0.05$, are displayed in Fig.~\ref{Potential} as red points, whereas the black points represent the Schwarzschild LR position. This spacetime exhibits the unusual characteristic of having a LR outside the equatorial plane, i.e. $\theta_\pm\neq\pi/2$. The absence of $\mathbb{Z}_2$ symmetry causes the Schwarzschild LR - which is the same for both circulation directions - to move in the $\theta$-direction either upwards or downwards, depending on the photon's circulation direction. Consequently, the Schwarzschild LR splits into two separate LRs: one positioned above the equatorial plane and another below it. A similar effect was reported to occur in the Taub-NUT solution~\cite{Wu:2023eml}, and Kerr BHs surrounded by plasma~\cite{Perlick:2023znh}.

\begin{figure}
	\centering
	\includegraphics[width=\columnwidth]{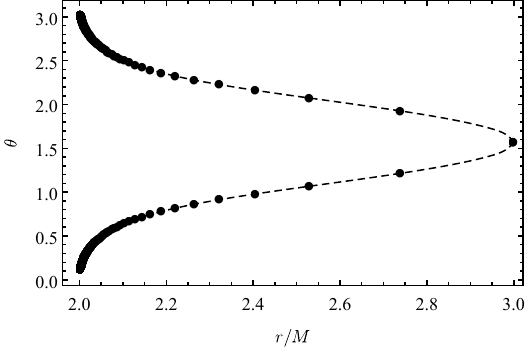}
	\caption{LR position on the $(r,\theta)$-plane for various values of the rotation parameter $j$. We considered a sequence of values of $j$ defined by $j=0.025\times n$ for $n\in\mathbb{N}$. The sequence starts at $j=0$, for which the LR is located at $(3M,\pi/2)$, and for $j\to\infty$ the LR positions go to $(2M,0)$ and $(2M,\pi)$ in the used coordinate system, which, albeit not geometrically invariant, clearly shows the off-equatorial displacement of the LRs with increasing $j$.}
	\label{LRj}
\end{figure}

In Fig.~\ref{LRj} we display the LR position as we increase the value of the background rotation parameter $j$. The point on the far right of Fig.~\ref{LRj} corresponds to the Schwarzschild LR. As we increase $j$, the LR split into two, one for each potential $H_{\pm}$, such that $\theta_+<\pi/2$ ($\theta_{-}>\pi/2$) is associated with $H_{+}$ ($H_{-}$). The points $(2M,0)$ and $(2M,\pi)$ in the $(r,\theta)$-space are accumulation points of the sequence presented in Fig.~\ref{LRj}.

\begin{figure}[h!]
	\centering
	\includegraphics[width=\columnwidth]{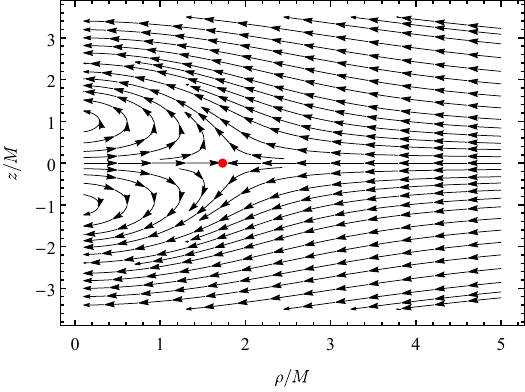}
	\includegraphics[width=\columnwidth]{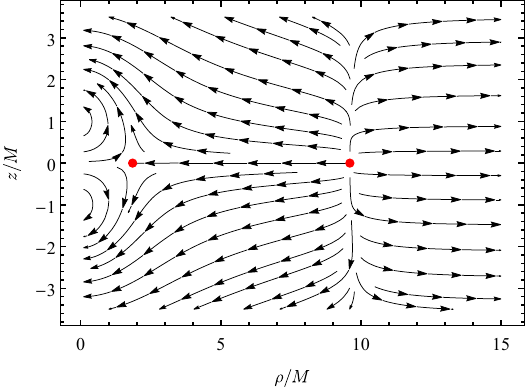}
	\includegraphics[width=\columnwidth]{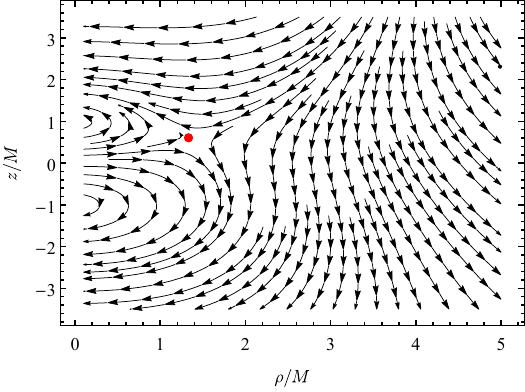}
	\caption{Vector field $\mathbf{v}_{+} = (v_{\rho}^+, v_z^+)$ in the $(\rho/M, z/M)$-plane for the Schwarzschild BH (top), Schwarzschild-Melvin BH with $B M=0.1$ (middle) and SBHSU with $jM^2 = 0.025$ (bottom). The red points represent the location of the LRs of the corresponding spacetime.}
	\label{VecPlot}
\end{figure}

The result presented in Fig.~\ref{LRj} is in accordance with Eq.~\eqref{Hsym}, which implies that if $(r,\theta)$ is a critical point of $H_{+}$, then $(r,\pi-\theta)$ must be a critical point of $H_{-}$. Hence, if $(r,\theta)$ corresponds to a LR position such that $\theta\neq\pi/2$, then there must exist another LR with the same radius, separated from the former by one $\theta$-reflection with respect to the equator. These two LRs also must share the same stability properties.

We remark that the correspondence between the labels $\pm$ of the potentials $H_{\pm}$ and the rotation sense requires careful analysis. In the case of Kerr, for instance, there exists two LRs that counter rotate in relation to each other. Due to the single direction of rotation in Kerr spacetime, one LR must co-rotate with the BH, while the other counter rotates. The SBHSU spacetime similarly accommodates two counter rotating LRs, but both of them co-rotate locally with the spacetime, as can be inferred from Eq.~\eqref{drag}.

A powerful technique to determine whether there are LRs in a given spacetime was developed in Ref.~\cite{Cunha:2020azh}. This technique consists of calculating a topological charge obtained from the circulation integral of the gradient of the potentials $H_{\pm}$. The sign of the topological charge defines the stability of the corresponding LR. It was shown that any stationary, axially symmetric, circular and asymptotically flat BH spacetime must have a topological charge equal to $-1$ for each $H_{\pm}$, indicating the existence of at least one unstable LR associated with each potential. 

Given that the SBHSU is not asymptotically flat, it falls outside the scope of the theorem proved in Ref.~\cite{Cunha:2020azh}. Our task now is therefore  to evaluate the topological charge for this case. We choose to work with the Weyl coordinate system defined by

\begin{equation}\label{W1}
	\rho=\sqrt{r^2-2Mr}\sin\theta,
\end{equation}
\begin{equation}\label{W2}
	z=(r-M)\cos\theta \ .
\end{equation}
The Weyl coordinates reduce to cylindrical coordinates in the asymptotic limit. In these coordinates, the horizon is located at $\rho=0$ and $|z|<M$. The axis of symmetry is determined by $\rho= 0$ and $|z|>M$. The remaining exterior region is defined by $\rho>0$. 

For any Jacobian matrix $\mathcal{J}$ connecting two coordinate systems $x^\mu$ and $x'^\mu$, it is true that, if both $\mathcal{J}\neq 0$ and $\mathcal{J}^{-1}\neq 0$, then the following equivalence holds: $\partial_i \phi(\bar{x}^\mu)=0\Leftrightarrow\partial'_i \phi(\bar{x}'^\mu)=0$, where $\phi$ is a scalar function. This means that $\bar{x}^\mu$ is a critical point of $\phi$ if and only if $\bar{x}'^\mu$ is a critical point of $\phi$. This condition is met for the coordinate transformation described by Eqs.~\eqref{W1} and~\eqref{W2}. Therefore, the task of identifying light rings (LRs) can be fully reformulated in terms of the Weyl coordinates.

The SBHSU metric in Weyl coordinates is given by
\begin{equation}
	\begin{aligned}\label{BHmetricW}
		&ds^2=\frac{  F(r,\theta )}{ f(r)+\left(M^2 \sin ^2\theta\right)/r^2}\left(d\rho^2+dz^2\right)\\-&f(r)F(r,\theta)dt^2+\frac{r^2\sin^2\theta}{F(r,\theta)}\Big\{d\varphi+\left[4jrf(r)\cos\theta\right]dt\Big\}^2,		
	\end{aligned}
\end{equation}
where the quantities $r$ and $\theta$ in Eq.~\eqref{BHmetricW} are to be understood as functions of $\rho$ and $z$, implicitly defined by Eqs.~\eqref{W1} and~\eqref{W2}. Let $\mathbf{v}_\pm$ be the vector fields defined by 
\begin{equation}
	\textbf{v}_{\pm}=\left(\frac{1}{\sqrt{g_{\rho\rho}}}\frac{\partial H_{\pm}}{\partial \rho}, \frac{1}{\sqrt{g_{zz}}}\frac{\partial H_{\pm}}{\partial z}\right)=\left(\text{v}_\rho^\pm,\text{v}_z^\pm\right).
\end{equation}

The explicit expression of these vector fields, even in terms of the coordinates $(r,\theta)$, is too lengthy and we choose not to show it here.  

We can write each component of the vector fields as
\begin{equation}\label{Omega1}
	\text{v}_\rho^\pm=|\textbf{v}_{\pm}|\cos\Omega(\rho,z),
\end{equation}
\begin{equation}\label{Omega2}
	\text{v}_z^\pm=|\textbf{v}_{\pm}|\sin\Omega(\rho,z),
\end{equation}
where $|\textbf{v}_{\pm}|=\sqrt{(\text{v}_\rho^\pm)^2+(\text{v}_z^\pm)^2}$ is the norm of $\textbf{v}_{\pm}$ and $\Omega$ is the angle formed by $\textbf{v}_{\pm}$ and the horizontal axis parametrized by $\text{v}_\rho^\pm$.

Let $\mathscr{C}$ be the homotopy class of (simple) curves in the $(\rho,z)$-space homotopic to the circle. The topological charge $w$ is a map $w:\mathscr{C}\to\mathbb{Z}$, defined to be
\begin{equation}\label{w}
	w_C=\frac{1}{2\pi}\oint_{C}d\Omega,
\end{equation}
where $C\in\mathscr{C}$. Eq.~\eqref{w} represents the winding number of the vector fields $\textbf{v}_{\pm}$ along some curve $C$ in the homotopy class $\mathscr{C}$ defined over the $(\rho,z)$-space. 

Each LR enclosed by $C$ contributes to the topological charge by $\pm 1$. Therefore, $w_C$ is a homotopic invariant if the number of LRs inside $C$ does not change. LRs with $w=-1$ are unstable, whereas LRs with $w=1$ are stable.

To calculate the total topological charge, one must consider a curve $C$, as defined in Eq.~\eqref{w}, that encompasses all the LRs. Hence, despite the complexity of the vector fields $\textbf{v}_{\pm}$ components in terms of the coordinates $(\rho, z)$, determining the topological charge only requires an analysis of its behavior in asymptotic regions, namely, near the horizon, the axis, and at infinity.

The vector plot of the $\textbf{v}_{+}=\{v^+_\rho,v^+_z\}$ field in terms of the Weyl coordinates can be seen in Fig.~\ref{VecPlot}. It suffices to restrict our analysis to  $\textbf{v}_{+}$, since we can always derive the conclusions for  $\textbf{v}_{-}$ by means of Eq.~\eqref{Hsym}.

In Fig.~\ref{VecPlot} we compare the vector field $\mathbf{v}_{+}$ of SBHSU with Schwarzschild spacetime and also with Schwarzschild BH immersed in the Melvin universe. The qualitative behavior of the vector field $\mathbf{v}_{+}$ near the horizon and close to the axis is similar for all the three cases. The primary distinction between Schwarzschild, Schwarzschild-Melvin and SBHSU  arises in the asymptotic region. The Melvin and SU asymptotics result in a positive asymptotic value for $v_\rho$, whereas it is negative for asymptotically flat spacetimes. Nevertheless, the vector fields $\mathbf{v}_{+}$ of Schwarzschild-Melvin and SBHSU are different for $z\gg 1$, namely, the vector fields turns in different directions to flip the sign of $v_\rho$. This difference will have a decisive impact on the topological charge of SBHSU, when compared to Schwarzschild-Melvin.

\begin{figure}[h!]
	\centering
	\includegraphics[width=\columnwidth]{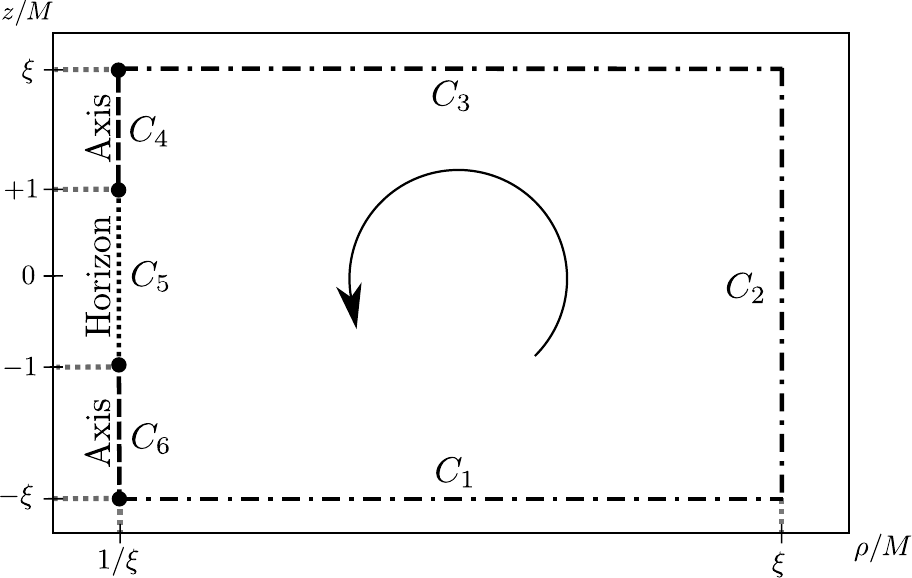}
	\caption{Schematic representation of the contour $C$, enclosing a compact region outside the horizon in the $(\rho,z)$-plane. The region inside the curve $C$ is designed to encompass all of the spacetime outside the horizon as $\xi\rightarrow \infty$.}
	\label{Diagram}
\end{figure}

Given a real number $\xi>0$, the total topological charge is computed by considering a curve $C=\bigcup_{k=1}^{6}C_k$. In this arrangement, the $z$ coordinate of the paths $C_1$ and $C_3$ is fixed at $-M\xi$ and $M\xi$, respectively. Meanwhile, for the other paths $C_2$ and the combination $C_4\cup C_5\cup C_6$, the radial coordinate $\rho$ is fixed at $M\xi$ and $M/\xi$, respectively (see Fig.~\ref{Diagram}).

Thus, we can separate the integral of Eq.~\eqref{w} into six pieces, according to
\begin{equation}
	2\pi w_C=\mathcal{I}_1+\mathcal{I}_2+\mathcal{I}_3+\mathcal{I}_4+\mathcal{I}_5+\mathcal{I}_6,
\end{equation}
where
\begin{equation}\label{I1}
	\mathcal{I}_1=\left[\int_{M/\xi}^{M\xi}\frac{d\Omega}{d\rho}d\rho\right]_{z=-M\xi},
\end{equation}
\begin{equation}\label{I2}
	\mathcal{I}_2=\left[\int_{-M \xi}^{M\xi}\frac{d\Omega}{dz}dz\right]_{\rho=M\xi},
\end{equation}
\begin{equation}\label{I3}
	\mathcal{I}_3=\left[\int_{M\xi}^{M/\xi}\frac{d\Omega}{d\rho}d\rho\right]_{z=M\xi},
\end{equation}
\begin{equation}\label{I4}
	\mathcal{I}_4=\left[\int_{M\xi}^{M}\frac{d\Omega}{dz}dz\right]_{\rho=M/\xi},
\end{equation}
\begin{equation}\label{I5}
	\mathcal{I}_5=\left[\int_{M}^{-M}\frac{d\Omega}{dz}dz\right]_{\rho=M/\xi},
\end{equation}
\begin{equation}\label{I6}
	\mathcal{I}_6=\left[\int_{-M}^{-M\xi}\frac{d\Omega}{dz}dz\right]_{\rho=M/\xi}.
\end{equation}

Hence, the total topological charge in the exterior region of the BH is given by the limit
\begin{equation}\label{TC}
	w=\lim\limits_{\xi\rightarrow\infty} w_C.
\end{equation}

The composite curve formed by $C_4\cup C_6$ represents the axis limit, while $C_5$ denotes the horizon limit. Additionally, the path formed by the combination $C_1\cup C_2\cup C_3$ corresponds to the asymptotic limit.

In order to compute the integrals presented in Eqs.~\eqref{I1}-\eqref{I6}, it is necessary to ensure that the associated angle $\Omega$, within each path $C_k$, falls within the valid range for the inverse trigonometric functions $\arccos$ and $\arcsin$. Thus, $\Omega$ must belong, along each path $C_k$, to one of the following four intervals: $(\pi/2, -\pi/2)$, $(\pi/2, 3\pi/2)$, $(0, \pi)$, or $(-\pi, 0)$. 

For each of the four possibilities, $\Omega$ is calculated inverting Eqs.~\eqref{Omega1} and~\eqref{Omega2} according to
\begin{equation}\label{O1}
	\text{if}\ \Omega\in(\pi/2, -\pi/2):\ \Omega=\arcsin\left(\frac{v_z^\pm}{|\textbf{v}_{\pm}|}\right) \ ,
\end{equation}
\begin{equation}\label{O2}
	\text{if}\ \Omega\in(\pi/2, 3\pi/2):\ \Omega=\pi-\arcsin\left(\frac{v_z^\pm}{|\textbf{v}_{\pm}|}\right) \ ,
\end{equation}
\begin{equation}\label{O3}
	\text{if}\ \Omega\in(0, \pi):\ \Omega=\arccos\left(\frac{v_\rho^\pm}{|\textbf{v}_{\pm}|}\right) \ ,
\end{equation}
\begin{equation}\label{O4}
	\text{if}\ \Omega\in(-\pi, 0):\ \Omega=-\arccos\left(\frac{v_\rho^\pm}{|\textbf{v}_{\pm}|}\right) \ .
\end{equation}

For any of the four possibilities, the integrals $\mathcal{I}_k$  are calculated by substituting $\Omega$ for the formulae given in Eqs.~\eqref{O1}-\eqref{O4}, evaluated at the corresponding integration limits.

\subsubsection{Asymptotic limit}
In the asymptotic limit we have to calculate the Integrals $\mathcal{I}_1$, $\mathcal{I}_2$ and $\mathcal{I}_3$. Starting with the integral $\mathcal{I}_1$, we have $z=-M\xi\rightarrow-\infty$, where the vector field $\mathbf{v}_{+}$ can be approximated as:

\begin{equation}
	v^+_\rho\approx-\frac{1-3 j^2 \rho ^4}{\rho ^2 \sqrt{1+j^2 \rho ^4}}+\mathcal{O}(\xi^{-1}),
\end{equation}
\begin{equation}
	v^+_z\approx-\frac{4 j}{\sqrt{1+j^2 \rho ^4}}+\mathcal{O}(\xi^{-1}),
\end{equation}
with higher-order terms represented by $\mathcal{O}(\xi^{-1})$. In this part of the integration, the corresponding angle $\Omega$ will be denoted by $\Omega_{z\to-\infty}$.

Since the $z$-component, $v^+_z$, is negative ($v^+_z<0$), it indicates that the angle $\Omega_{z\to-\infty}\in(-\pi,0)$, which implies that the function $\Omega_{z\to-\infty}$ can be calculated via Eq.~\eqref{O4}. Thus, the expression for the first integral can be written as
\begin{equation}
	\begin{aligned}
		\mathcal{I}_1&=\pi-\frac{1}{3 \sqrt{2}}\frac{1}{\xi}+\mathcal{O}(\xi^{-2}).
	\end{aligned}
\end{equation}

Now we may consider the integration for $\rho=M\xi\to\infty$, where the vector field $\mathbf{v}_{+}$ is approximately given by

\begin{equation}
	v^+_\rho\approx 3 j+\mathcal{O}(\xi^{-1}),
\end{equation}
\begin{equation}
	v^+_z\approx0+\mathcal{O}(\xi^{-1}).
\end{equation}

Since, up to zeroth order in $\xi^{-1}$, the vector field $\textbf{v}_{+}$ does not depend on neither of the coordinates $\rho$ and $z$, as $\rho$ approaches infinity, we can deduce that the contribution $\mathcal{I}_2$ from path $C_2$ to the topological charge is zero:
\begin{equation}
	\begin{aligned}
		\mathcal{I}_2=0 \ .
	\end{aligned}
\end{equation}

To complete the asymptotic analysis, we have to calculate $\mathcal{I}_3$. In this integration we let $z=M\xi\to\infty$, where the vector field $\mathbf{v}_{+}$ is approximately given by

\begin{equation}
	v^+_\rho\approx-\frac{1-3 j^2 \rho ^4}{\rho ^2 \sqrt{1+j^2 \rho ^4}}+\mathcal{O}(\xi^{-1}),
\end{equation}
\begin{equation}
	v^+_z\approx-\frac{4 j}{\sqrt{1+j^2 \rho ^4}}+\mathcal{O}(\xi^{-1}).
\end{equation}

Thus, the calculation is performed exactly the same as it was for the $C_1$ path, just swapping the integration limits, which gives

\begin{equation}
	\mathcal{I}_3=-\pi+\frac{1}{3 \sqrt{2}} \frac{1}{\xi}+\mathcal{O}(\xi^{-2}).
\end{equation}

\subsubsection{Axis limit}
 In this part of integration we let $\rho=M/\xi\to0$. The coordinate $z$ has to satisfy either $z>M$ or $z<-M$. We first consider the case $z>M$, which is associated with the $\mathcal{I}_4$ integral. Thus, we can make an approximation for the vector field $\mathbf{v}_{+}$ as follows:
\begin{equation}
	v^+_\rho\approx-\left(\frac{z-M}{z+M}\right)^{3/2}\frac{\xi^2}{M^2}+\mathcal{O}(\xi^0) \ ,
\end{equation} 
\begin{equation}
	v^+_z\approx\frac{2   \sqrt{z-M}}{(z+M)^{5/2}}\xi+\mathcal{O}(\xi^{0}) \ .
\end{equation} 
The corresponding angle $\Omega$ will be denoted by $\Omega_{\rho\to0}^{z>M}$. Since the $\rho$-component, $v^+_\rho$, is negative ($v^+_\rho>0$), it indicates that the angle $\Omega_{\rho\to0}^{z>M}$ lies between $\pi/2$ and $3\pi/2$, which implies that Eq.~\eqref{O2} is suitable for this calculation. Thus, the expression for $\mathcal{I}_4$ can be written as
\begin{equation}
	\mathcal{I}_4=-\frac{\pi }{2}+\frac{8 j M^2}{\xi }+\mathcal{O}(\xi^{-2}).
\end{equation}

Analogously, we may calculate $\mathcal{I}_6$ considering $z<-M$. We obtain that the vector field $\mathbf{v}_{+}$ takes the approximated form:
\begin{equation}
	v^+_\rho\approx-\left(\frac{z+M}{z-M}\right)^{3/2}\frac{\xi^2}{M^2}+\mathcal{O}(\xi^0) \ ,
\end{equation} 
\begin{equation}
	v^+_z\approx-\frac{2   \sqrt{-z-M}}{(-z+M)^{5/2}}\xi+\mathcal{O}(\xi^{0}) \ .
\end{equation} 

The corresponding angle $\Omega$ will be denoted by $\Omega_{\rho\to0}^{z<-M}$. Since the $\rho$-component, $v^+_\rho$, is negative ($v^+_\rho<0$), it indicates that the angle $\Omega_{\rho\to0}^{z<-M}$ lies between $\pi/2$ and $3\pi/2$, which implies that we must use Eq.~\eqref{O2}. Thus, the expression for $\mathcal{I}_6$ is given by

\begin{equation}
	\mathcal{I}_6=-\frac{\pi }{2}+\frac{8 j M^2}{\xi }+\mathcal{O}(\xi^{-2}).
\end{equation}

\subsubsection{Horizon limit}
Following the $C_5$ curve, we still have $\rho=M/\xi\to0$, but now $|z|<M$. We want to show that $v_\rho^+>0$, hence we may use Eq.~\eqref{O1} to evaluate $\mathcal{I}_5$. In order to do so, we have to go back do spherical coordinates. Thus, we have that
\begin{equation}
	v_\rho^+=\frac{\partial_\rho H_+}{\sqrt{g_{\rho \rho}}}=\frac{\left(\frac{\partial r}{\partial \rho}\frac{\partial}{\partial r}+\frac{\partial \theta}{\partial \rho}\frac{\partial}{\partial \theta}\right)H_+}{\sqrt{\left(\frac{\partial r}{\partial \rho }\right)^2g_{rr}+\left(\frac{\partial \theta }{\partial \rho }\right)^2g_{\theta \theta }}}\underset{r\to 2M}{\approx}\frac{\partial_r H_+}{\sqrt{g_{rr}}}.
\end{equation}

Therefore, the $\rho$-component of the vector field $\mathbf{v}_{+}$, nearby the horizon ($r=2M+\delta r$, where $\delta r$ is a small, positive real number), in spherical coordinates, can be written as
\begin{equation}
	v_\rho^+\approx\frac{\sqrt{F(2M,\theta)}}{8 M^2 \sin \theta}-\sqrt{\frac{2}{M}}\frac{2j\cos \theta}{\sqrt{F(2M,\theta)}} \delta r^{1/2}+\mathcal{O}(\delta r).
\end{equation}

Since $\delta r\to0$, we have that $v_\rho^+>0$. Thus, let $\Omega_{\rho\to0}^{|z|<M}$ denote the vector field angle with the horizontal axis. We must have $\Omega_{\rho\to0}^{|z|<M}\in(-\pi/2,\pi/2)$, implying that we may use Eq.~\eqref{O1}. The corresponding integral $\mathcal{I}_5$ must be given by
\begin{equation}
	\mathcal{I}_5=-\pi+\frac{16 j M}{\xi }+\mathcal{O}(\xi^{-2}).
\end{equation}

We can now compute the total topological charge using Eq.~\eqref{TC}, which is given by
\begin{equation}
	\begin{aligned}
		w&=\lim\limits_{\xi\to\infty}\frac{1}{2\pi}\left(\mathcal{I}_1+\mathcal{I}_2+\mathcal{I}_3+\mathcal{I}_4+\mathcal{I}_5+\mathcal{I}_6\right)\\
		&=\lim\limits_{\xi\to\infty}\left[-\pi+\frac{16 j M^2}{\xi }+\mathcal{O}(\xi^{-2})\right]=-\pi.
	\end{aligned}
\end{equation}
Therefore, we proved that there must exist one unstable LR, for all values of $j\in(0,\infty)$, associated with the vector field $\textbf{v}_{+}$. Due to the odd $\mathbb{Z}_2$ symmetry (see Appendix~\ref{Appendix}) manifested in Eq.~\eqref{Hsym}, we can also conclude that there is also another LR associated with $\textbf{v}_{-}$, with the same radial coordinate $\rho$ and reflected $z$-coordinate  with respect to the equatorial plane. Therefore, there are two LRs, one for each potential.

To conclude our analysis in this section concerning LRs, we turn our attention to the examination of the topological charge within the context of the background spacetime alone, excluding the presence of the Schwarzschild BH. To study the SU we have to consider the same calculation, but now setting $M=0$. For $M=0$, the formulae undergo a substantial simplification. The Weyl coordinates reduce to cylindrical coordinates (see Eq.~\eqref{cylindrical}).

The potentials $H_\pm$ are given by
\begin{equation}
	H_\pm=-4 j z\pm\left(j^2 \rho ^3+\frac{1}{\rho }\right).
\end{equation}

\begin{figure}[h!]
	\centering
	\includegraphics[width=\columnwidth]{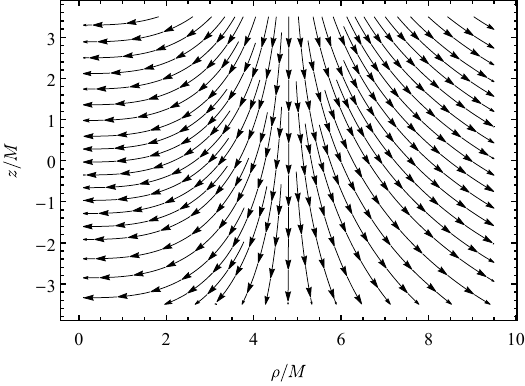}
	\caption{Plot illustrating the vector field $\mathbf{v}_{+} = (v_{\rho}^+, v_z^+)$ in the $(\rho/M, z/M)$-plane for the SU with $j = 0.025$. Since Eq.~\eqref{v0} does not depend on the coordinate $z$, the pattern of the vector field displayed in the plot repeats itself infinitely upwards and downwards.}
	\label{VecPlot00}
\end{figure}

The associated vector fields $\textbf{v}_{\pm}$ can be written as
\begin{equation}\label{v0}
	\textbf{v}_{\pm}=\left(\mp\frac{1-3 j^2 \rho ^4}{\rho ^2 \sqrt{1+j^2 \rho ^4}},-\frac{4 j}{\sqrt{1+j^2 \rho ^4}}\right).
\end{equation}
The vector fields $\textbf{v}_{\pm}$ of the SU are precisely the ones that we obtained for the SBHSU case in the limit $z\to\pm\infty$.

In Fig.~\ref{VecPlot00} we display a vector plot of $\textbf{v}_{+}$. Since the vector field does not depend on the $z$ coordinate, the winding number is zero for any curve $C$ in the $(\rho,z)$-plane. Therefore, the total topological charge is also zero, implying that in the SU there are no LRs.

To conclude, the LR of the SBHSU spacetime is actually the LR inherited from the Schwarzschild BH that was split and pushed off the equator by the  the swirling background.

\section{Shadows and gravitational lensing}
\label{Sec. IV}

\begin{figure}[h!]
	\centering
	\includegraphics[width=\columnwidth]{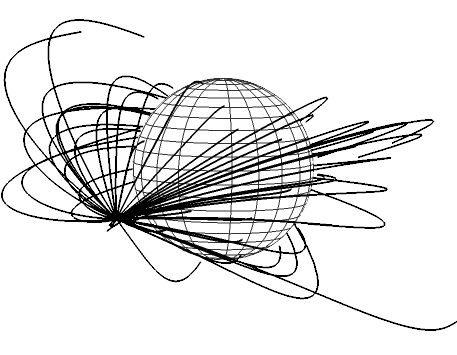}
	\caption{Plot illustrating the backward ray tracing method for the SBHSU spacetime. We evolved, numerically, 50 null geodesics with initial conditions given by Eqs.~\eqref{IC1}-\eqref{IC4}, $\theta=\pi/2$ and $r=10M$. We have chosen the observation angles at random.}
	\label{BRT}
\end{figure}

\begin{figure*}
	\centering
	\includegraphics[scale=0.22]{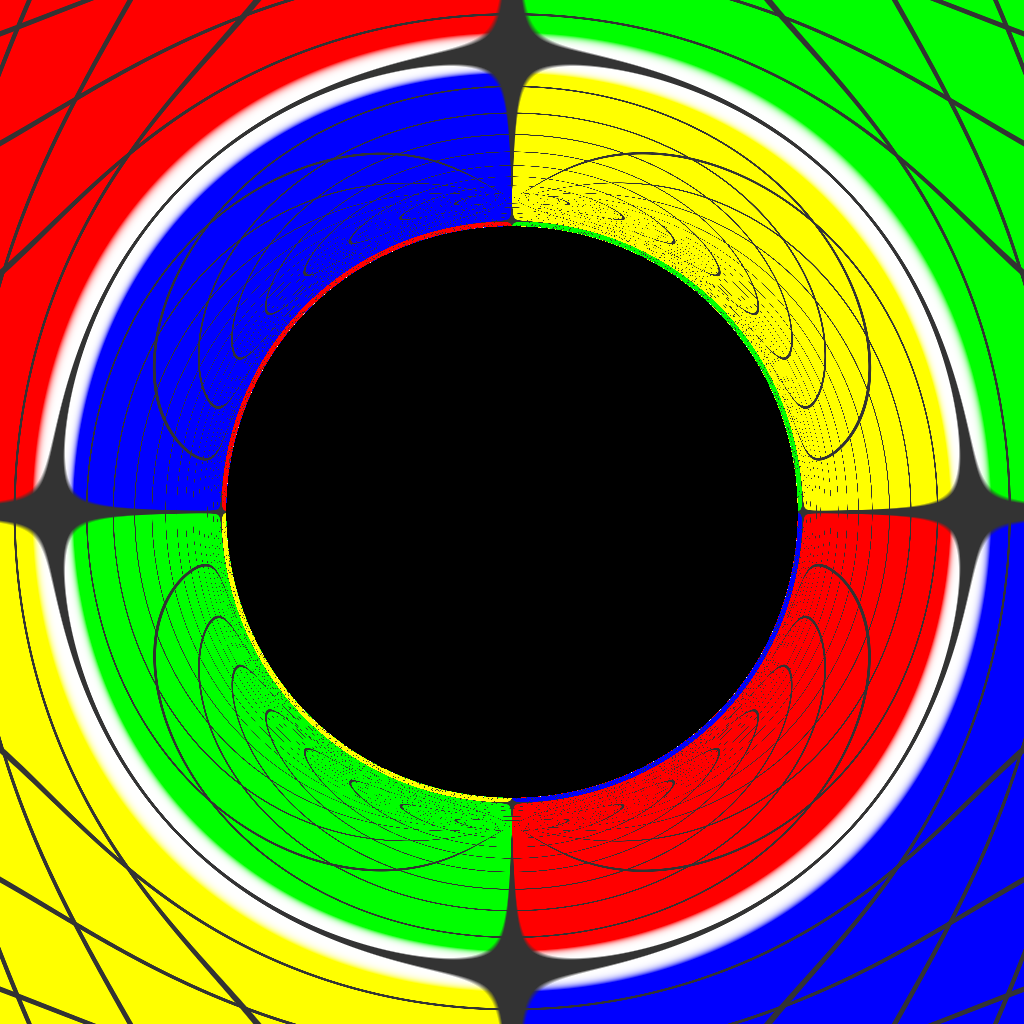}
	\includegraphics[scale=0.22]{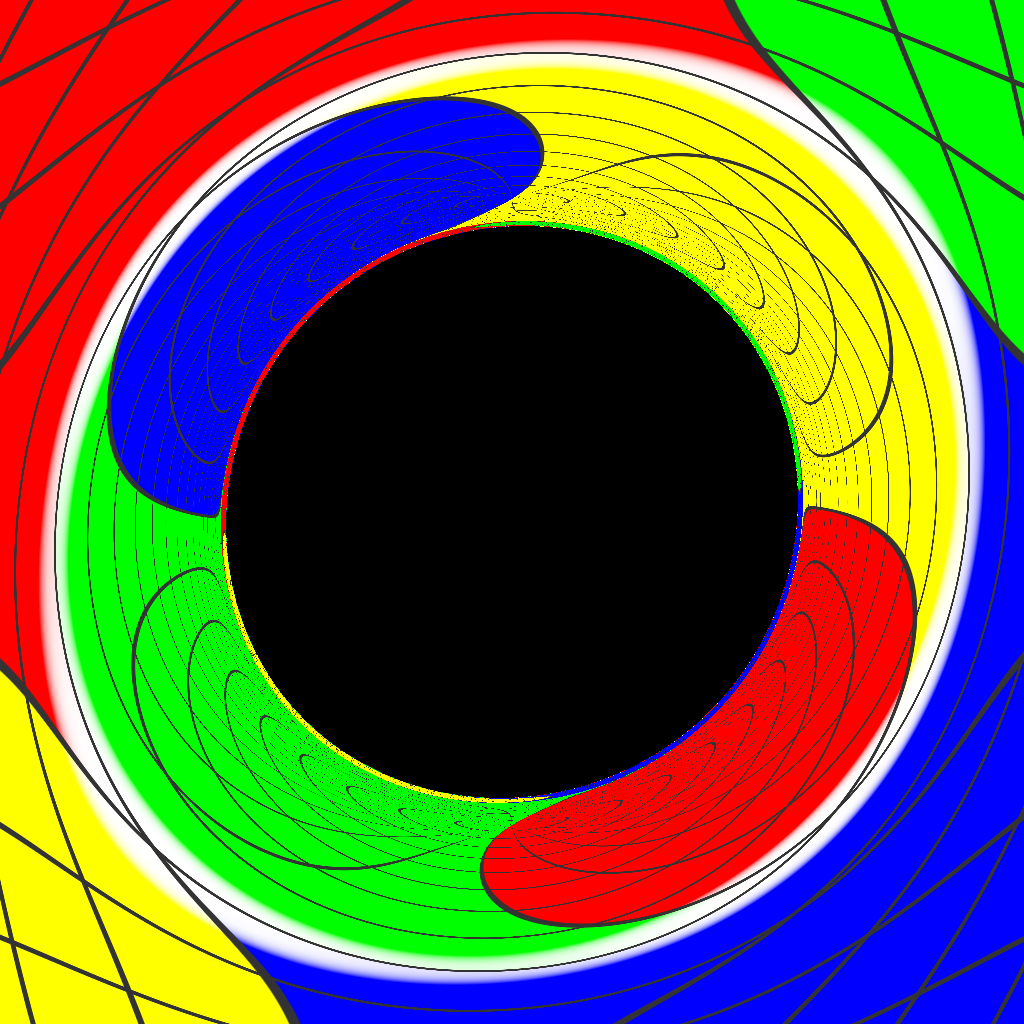}
	\includegraphics[scale=0.22]{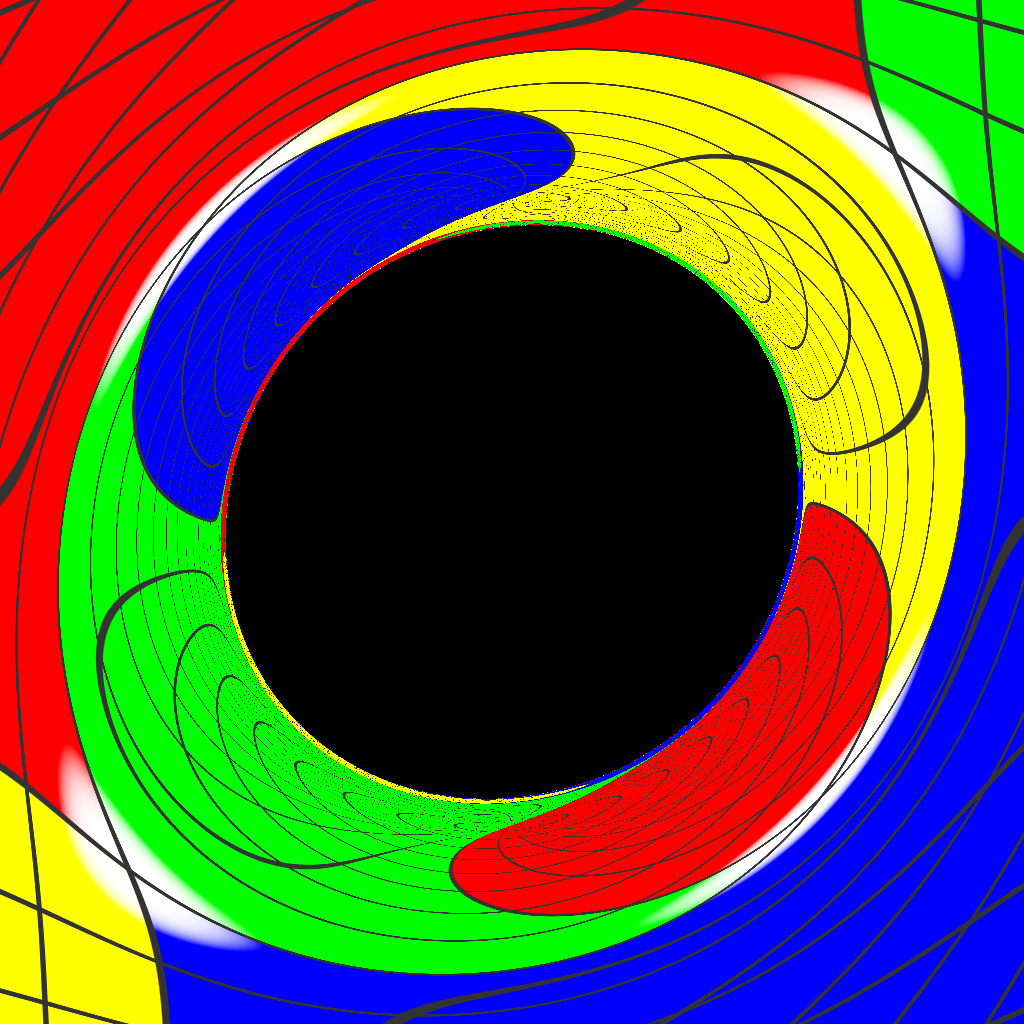}
	\includegraphics[scale=0.22]{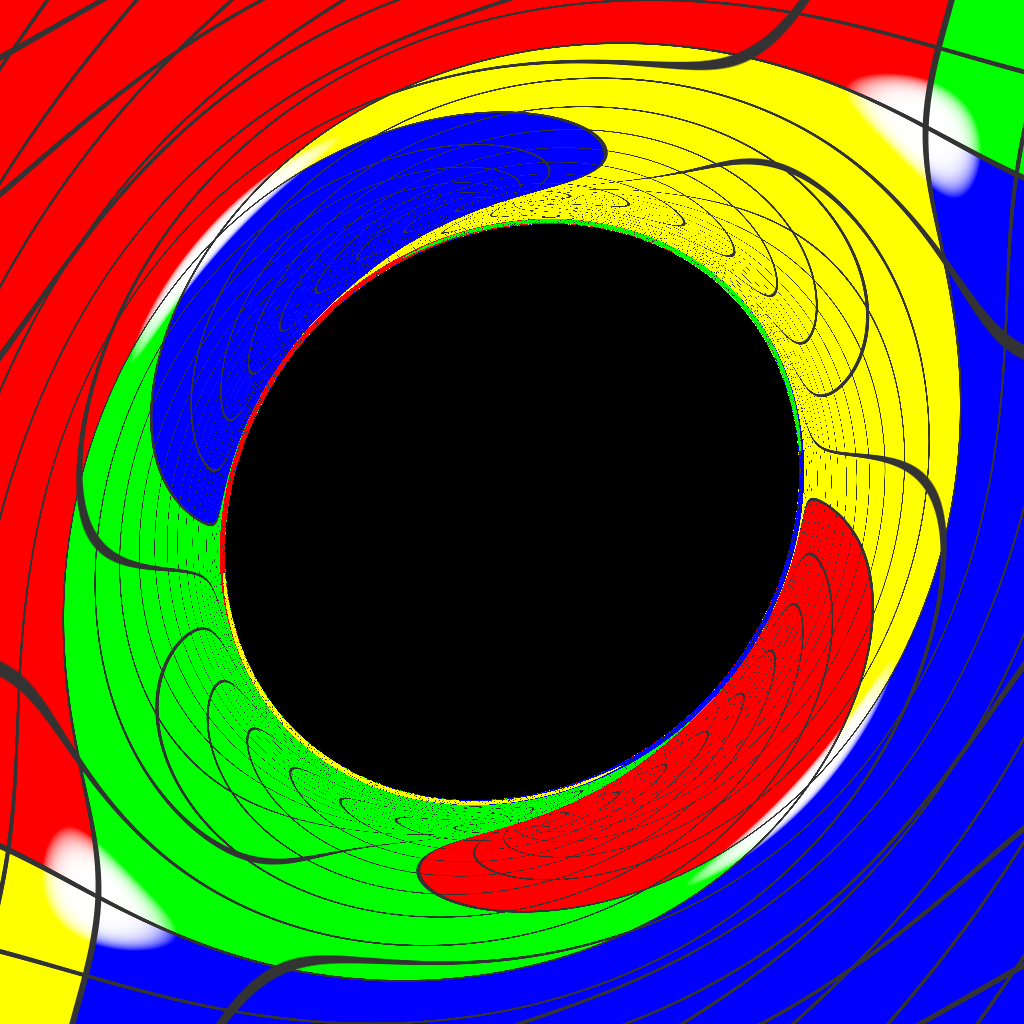}
	\includegraphics[scale=0.22]{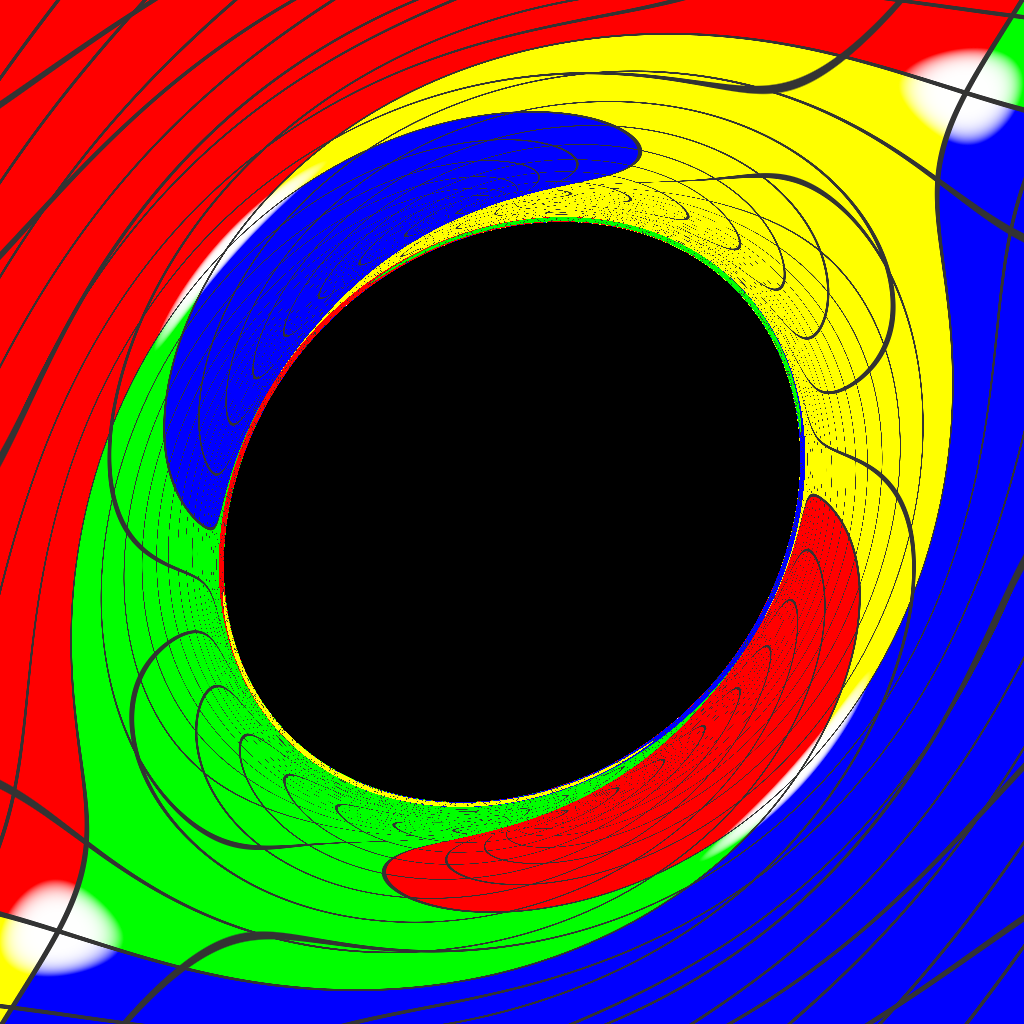}
	\includegraphics[scale=0.22]{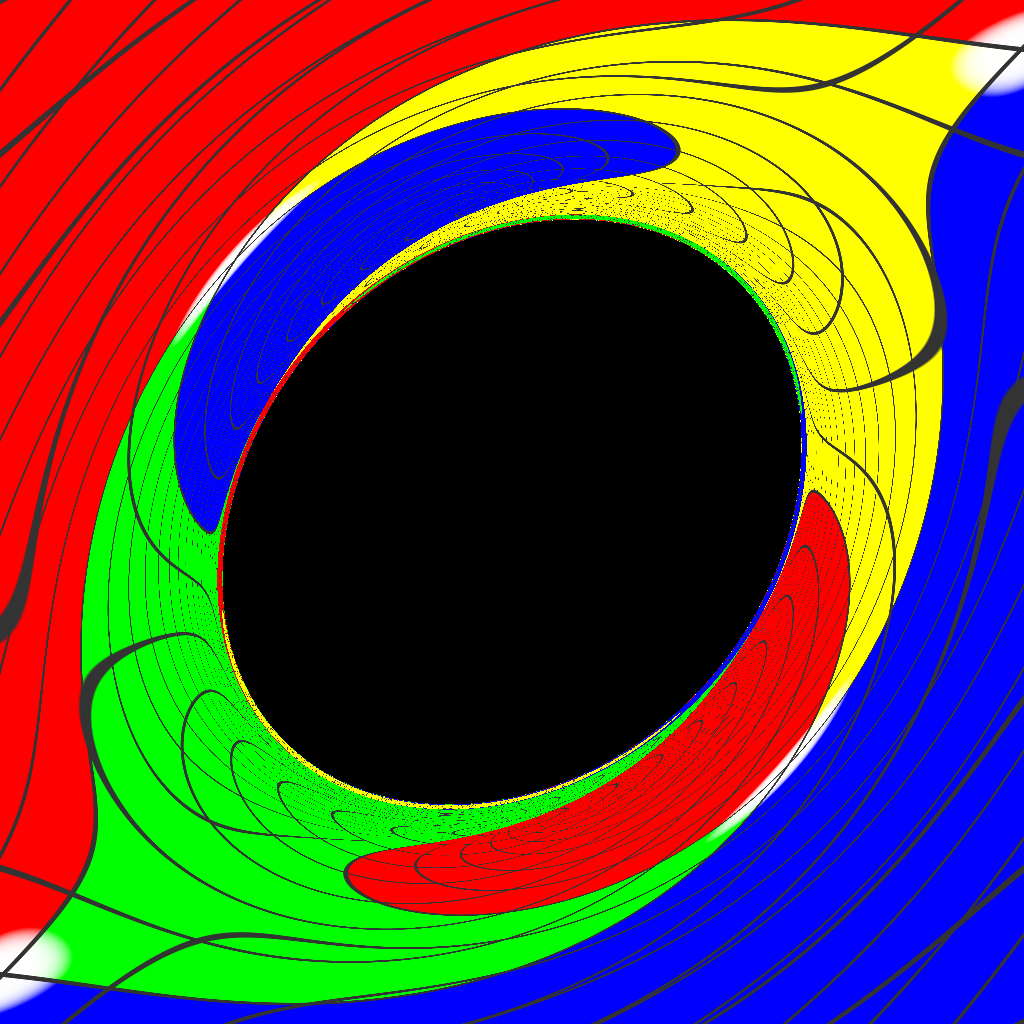}
	\caption{Shadow and gravitational lensing of the SBHSU, starting with the Schwarzschild case and increasing the swirling parameter $j$ in equal steps of $\delta j = 0.0001$. In all the images, the observer is
		positioned at the equatorial plane ($\theta=\pi/2$) and at the radius $r = 15M$.}
	\label{Shadow}
\end{figure*}

\begin{figure*}
	\centering
	\includegraphics[scale=0.22]{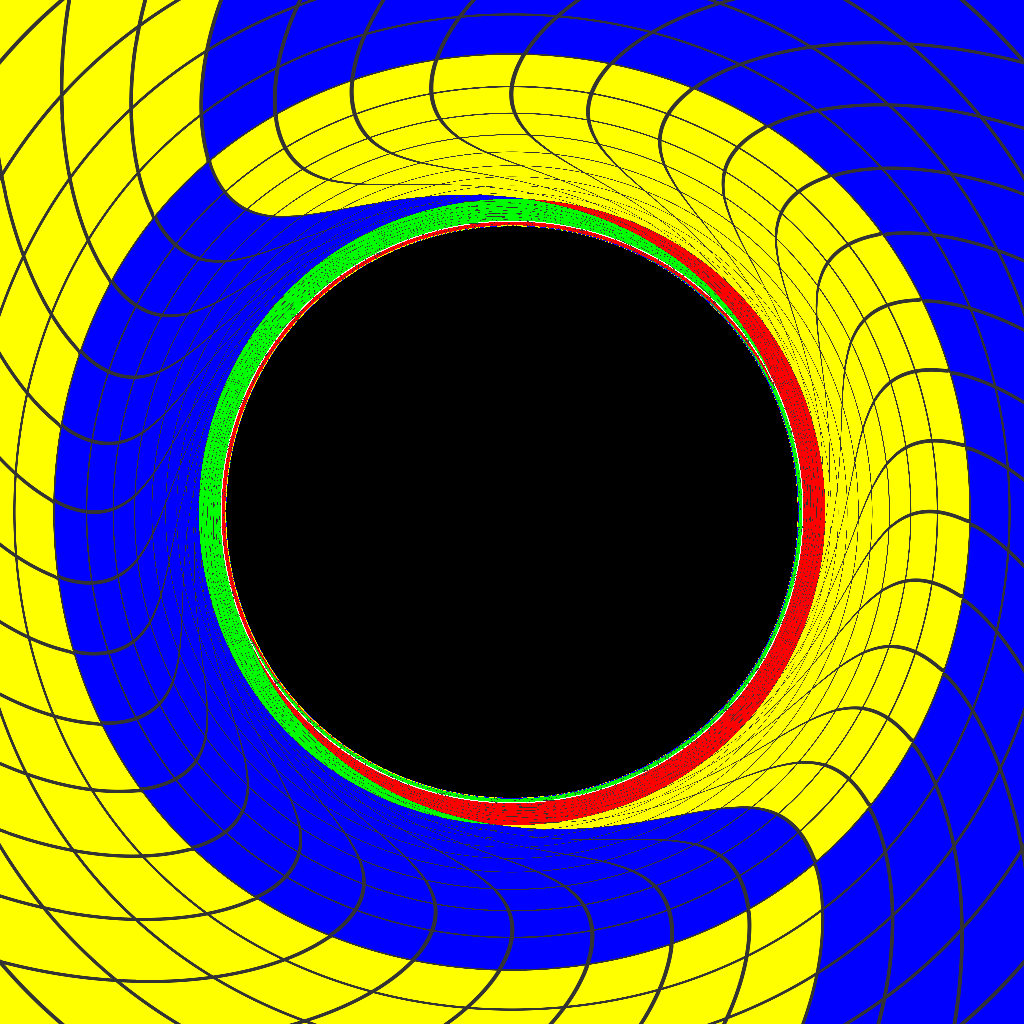}
	\includegraphics[scale=0.22]{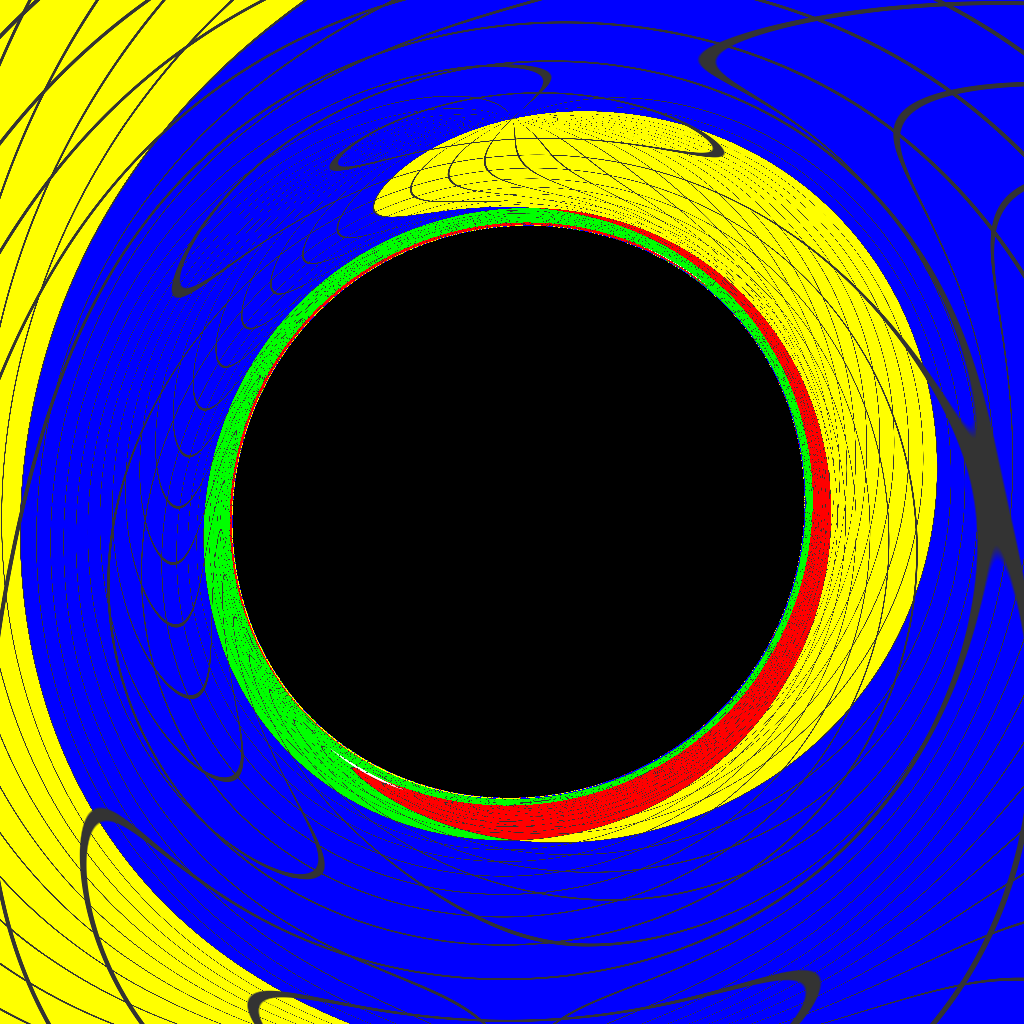}
	\includegraphics[scale=0.22]{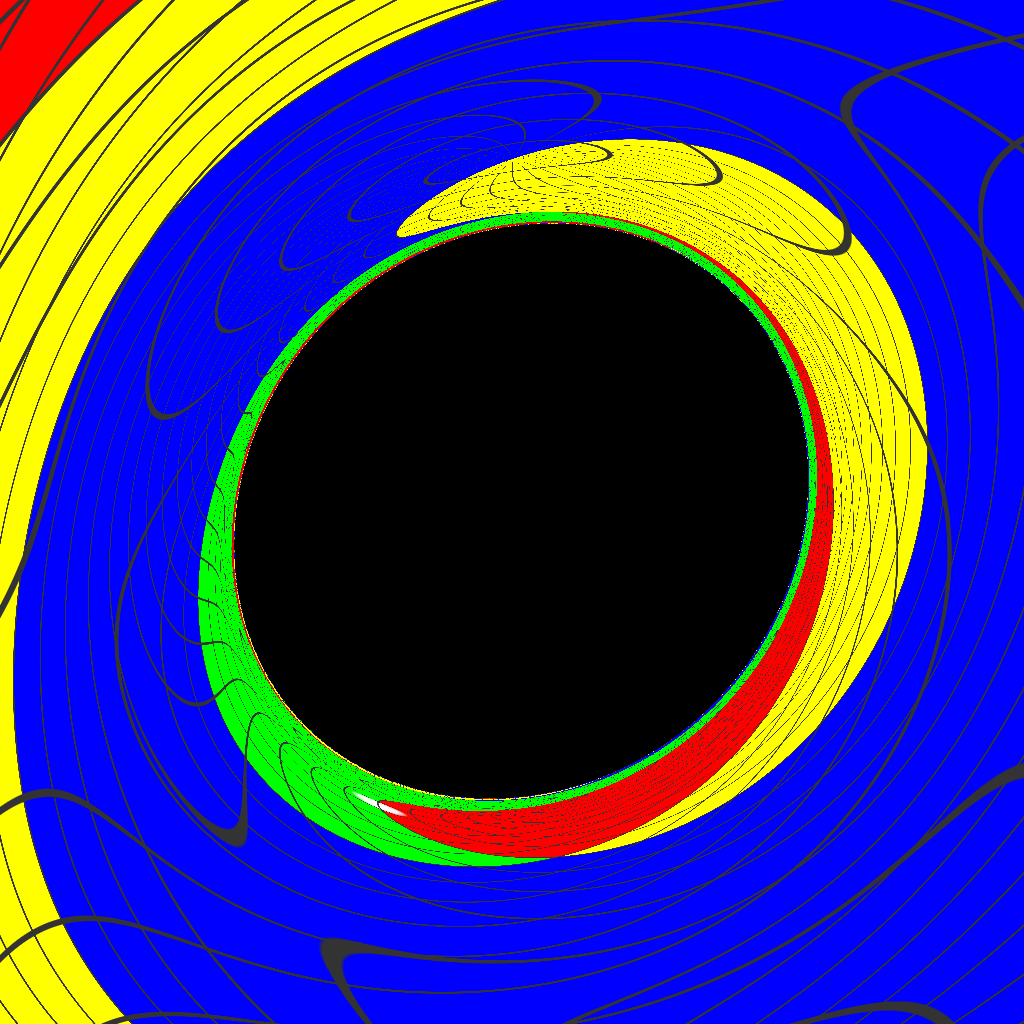}
	\includegraphics[scale=0.22]{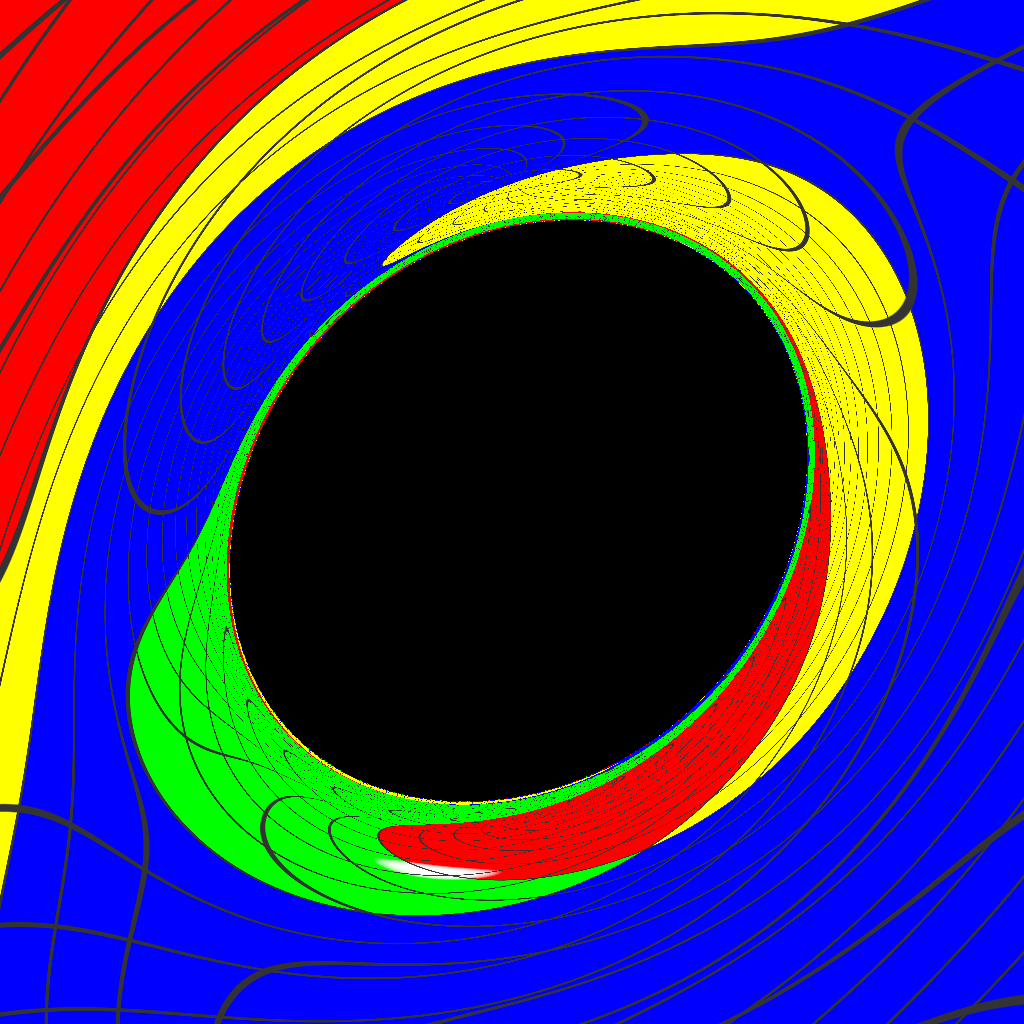}
	\includegraphics[scale=0.22]{1024NewSwirling0d0005}
	\includegraphics[scale=0.22]{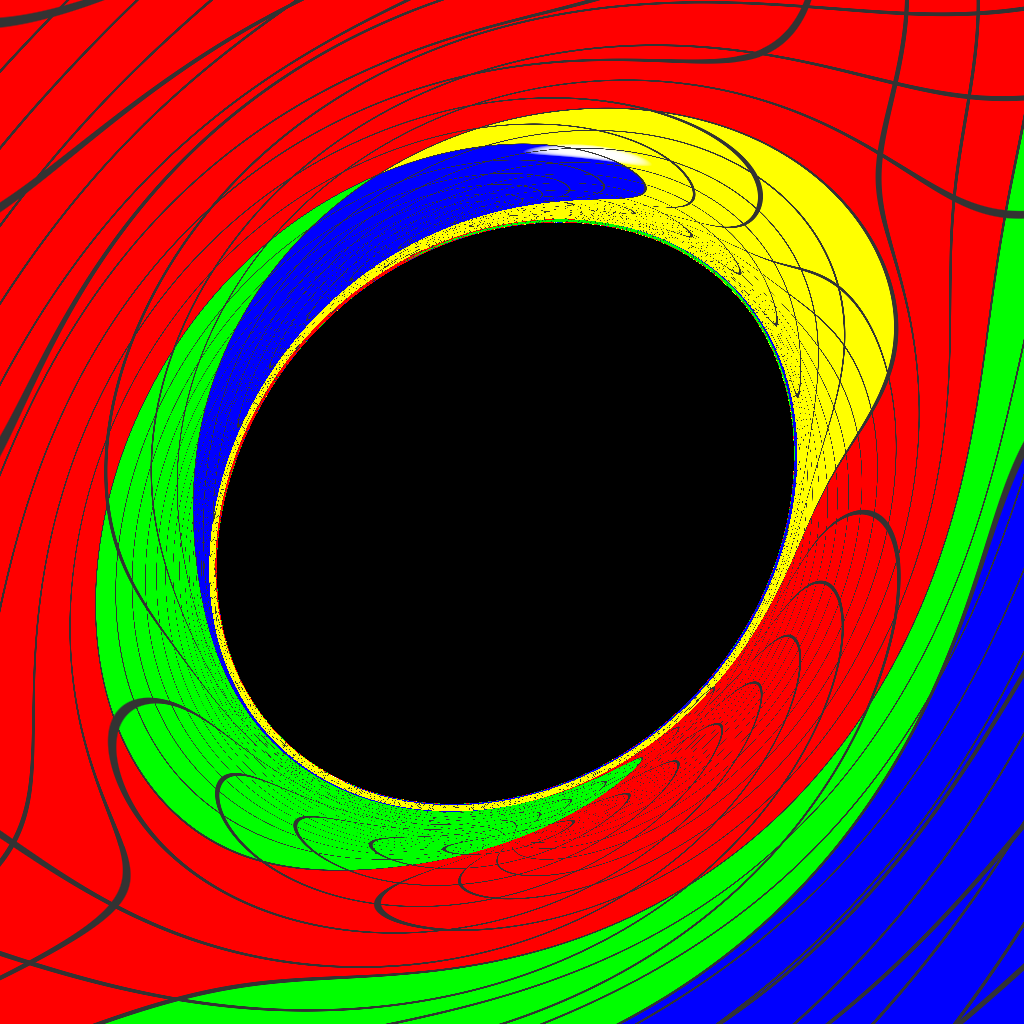}
	\includegraphics[scale=0.22]{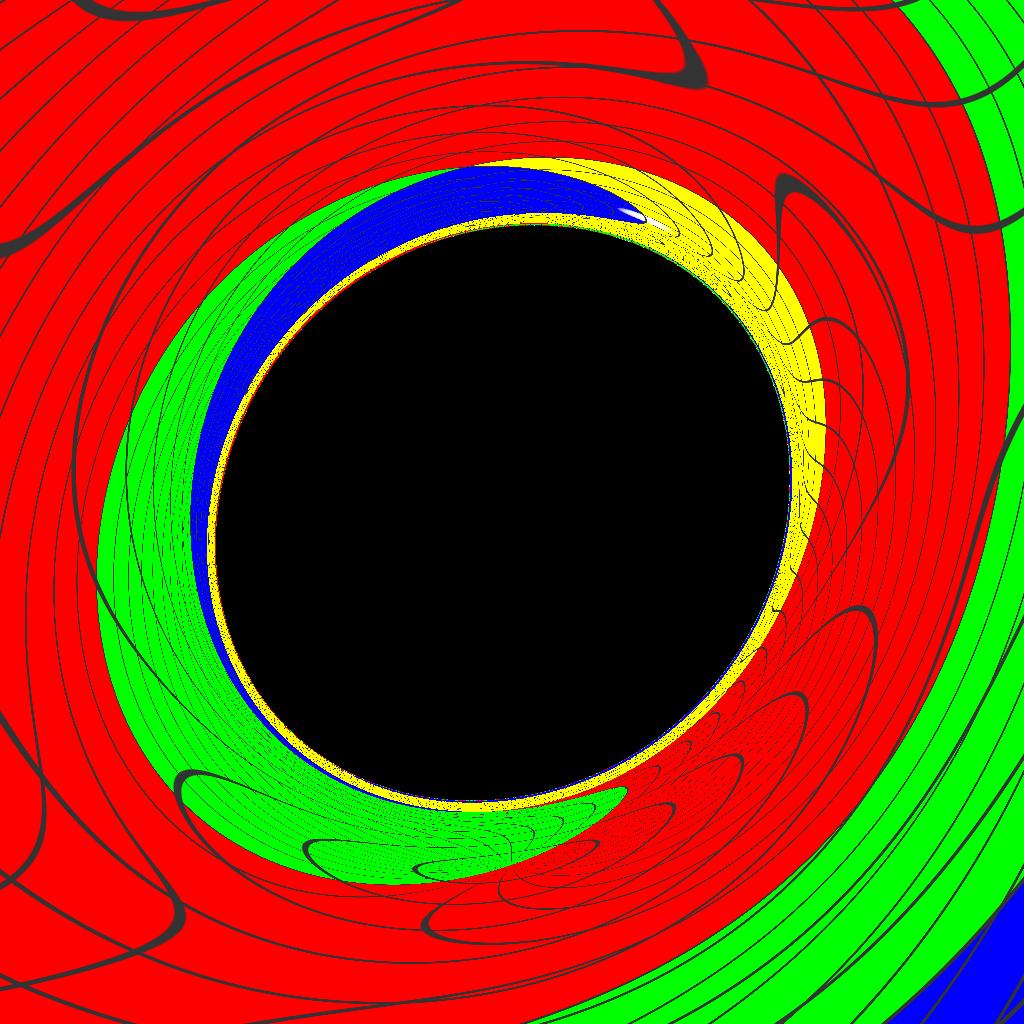}
	\includegraphics[scale=0.22]{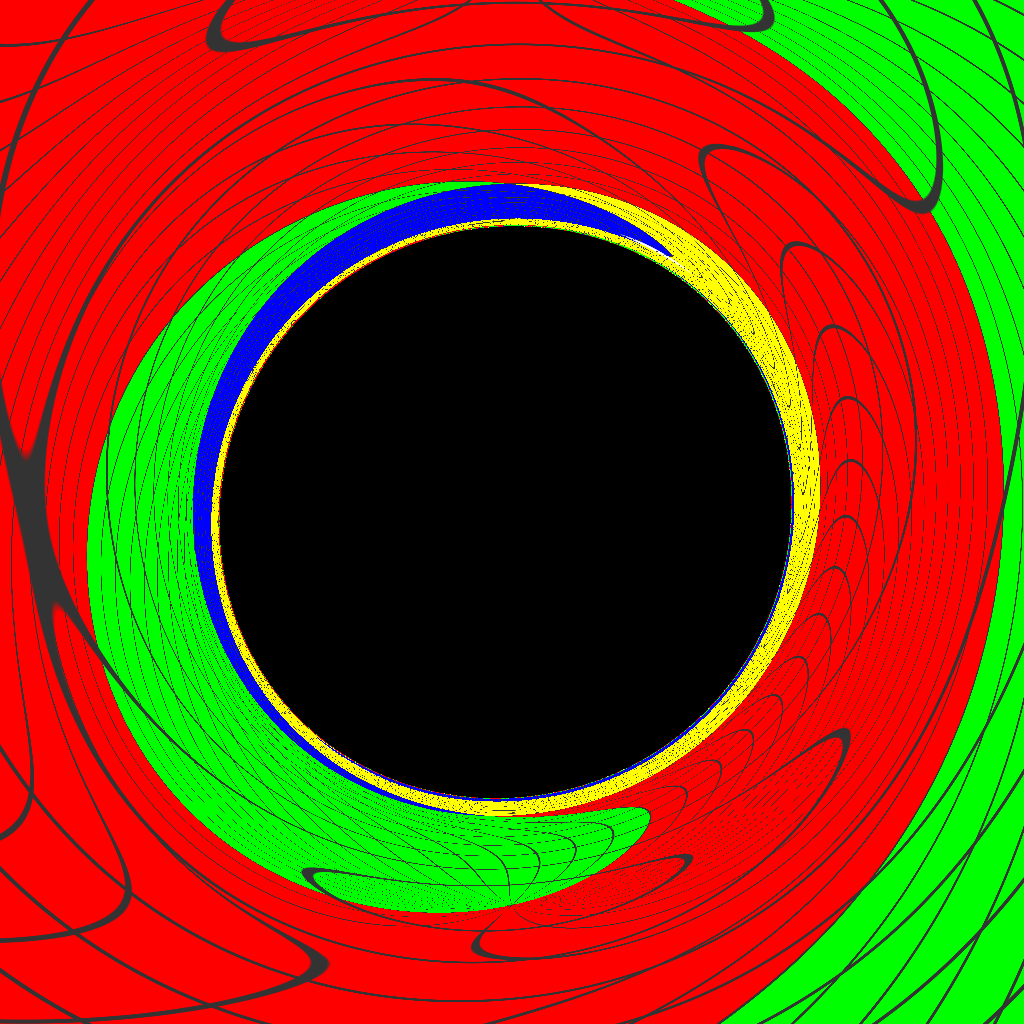}
	\includegraphics[scale=0.22]{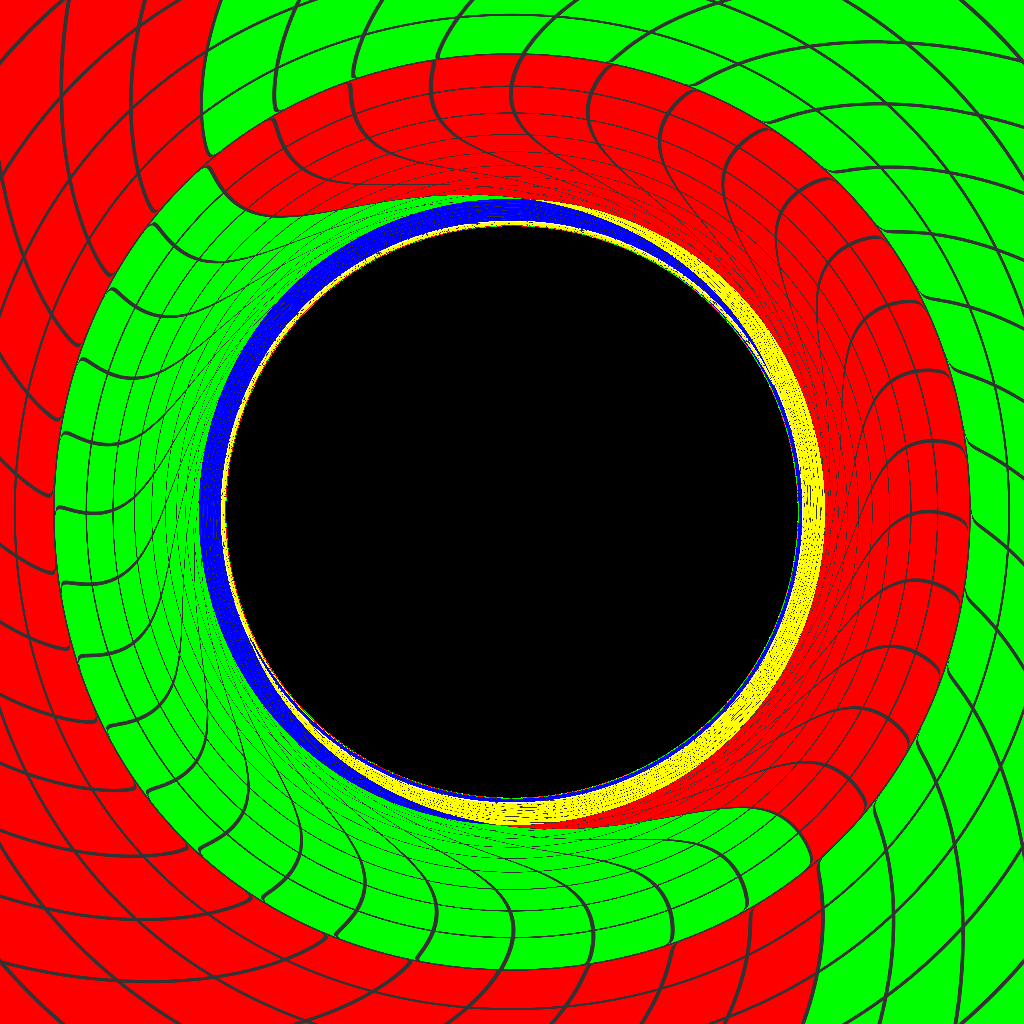}
	\caption{Shadow and gravitational lensing of the SBHSU, for $j M^2=0.0005$, for several observer positions. We start with $\theta=0$ (on axis observer seeing the BH from the top) and we increase the angle $\theta$ in equal steps of $\delta \theta = \pi/8$, until we reach $\theta=\pi$ (on axis observer seeing the BH from the bottom). We kept the radius fixed at $r = 15M$.}
	\label{Shadow2}
\end{figure*}

In this section we investigate the shadow and gravitational lensing phenomena of the SBHSU solution defined by Eq.~\eqref{BHmetric}. An analytic calculation of a BH shadow's edge depends on the integrability of the equations of motion of photons. Liouville integrable spacetimes  allow separation of variables and analytic expressions for the shadow's boundary. This is the case of Kerr BHs, which includes a Carter constant in addition to the conserved energy, angular momentum and particle's mass~\cite{Carter:1968rr}.

\begin{figure*}
	\centering
	\includegraphics[scale=0.5]{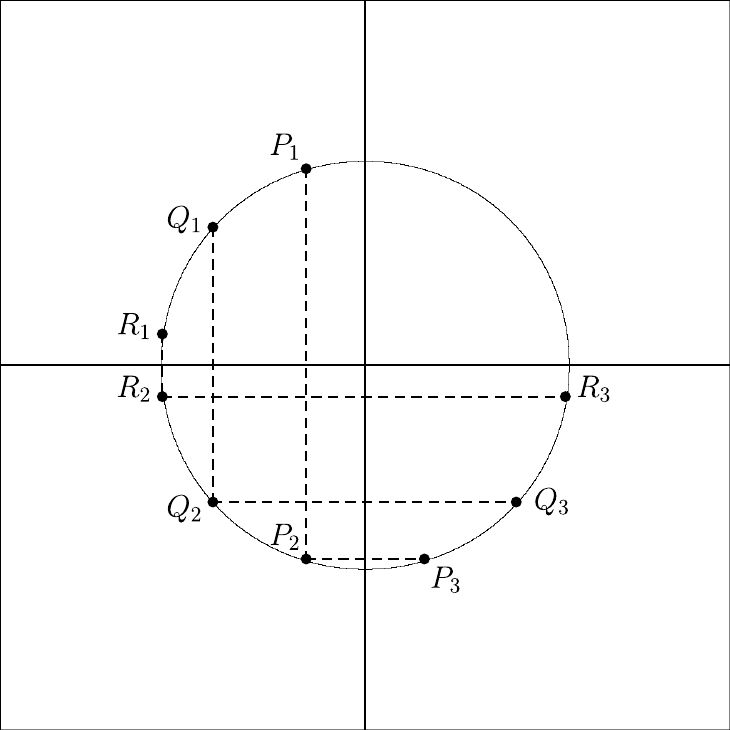}
	\includegraphics[scale=0.5]{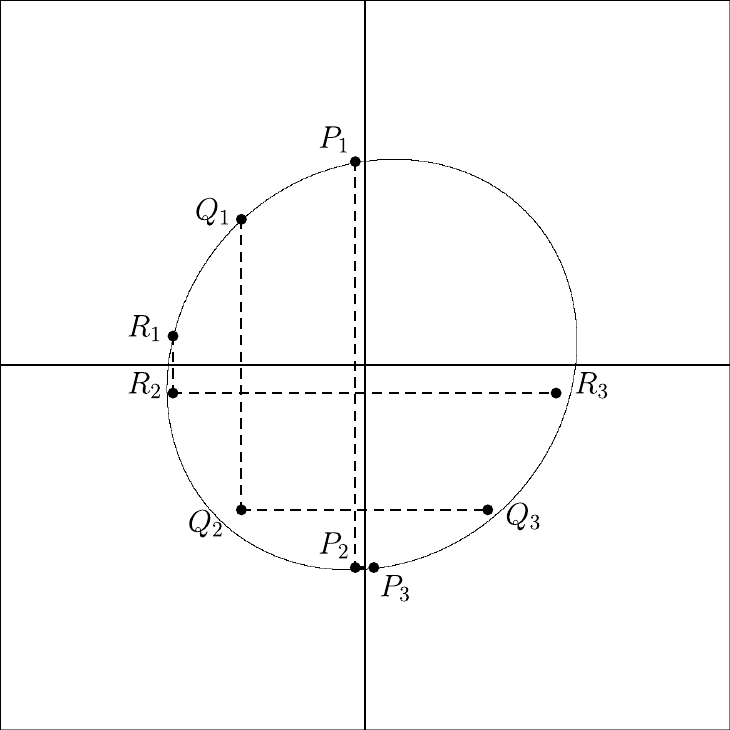}
	\includegraphics[scale=0.5]{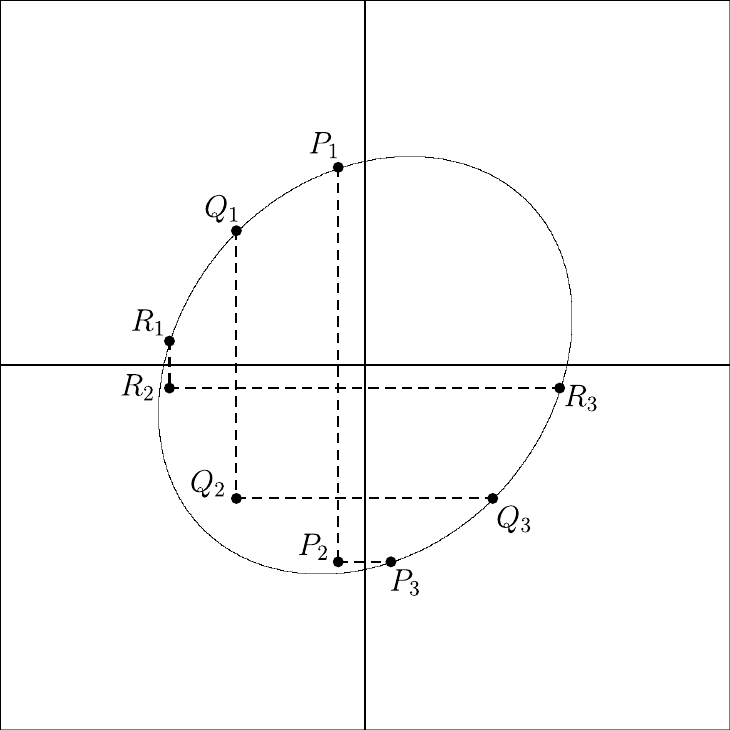}
	\caption{Shadow edge of the SBHSU for $j M^2=0.0005$, for observers positioned at $\theta=0$ (top left), $\theta=\pi/4$ (top right) and $\theta=\pi/2$ (bottom). We kept the radius fixed at $r = 15M$. The points $P_1$, $Q_1$ and $R_1$ are randomly chosen points on the shadow's edge. The points $P_2$, $Q_2$ and $R_2$ are obtained by reflecting  $P_1$, $Q_1$ and $R_1$ with respect to the horizontal axis at the middle of each plot. The points $P_3$, $Q_3$ and $R_3$ are obtained by reflecting  $P_2$, $Q_2$ and $R_2$ with respect to the vertical axis.}
	\label{Shadow3}
\end{figure*}

However, the SBHSU does not appear to be integrable in the $(t,r,\theta,\varphi)$ coordinates, in contrast with the SU background~\cite{Capobianco:2023kse}. In nonintegrable cases, the geodesic equations do not allow variable separability, leading, possibly, to chaotic photon motion. In those cases, a straightforward analytical shadow calculation is no longer possible. But one may calculate the shadow  by means of numerical simulations using (backward) ray tracing codes. The backward ray tracing technique consists of numerically evolving null geodesics backwards in time, starting from the observer's position, until the light rays either get absorbed by the BH or scattered to a celestial sphere. This is equivalent to consider the celestial sphere as a source of light, emitting radiation isotropically.

Here, we numerically integrate Eqs.~\eqref{xdot} and~\eqref{pdot}, with initial conditions given by
\begin{equation}\label{IC1}
	E=\sqrt{f(r) F(r,\theta )}-\frac{4 \mathit{j} r^2 \cos \alpha \sin\beta f(r) \sin\theta \cos\theta}{\sqrt{F(r,\theta )}},
\end{equation}
\begin{equation}\label{IC2}
	p_r=\cos \alpha \cos\beta \sqrt{\frac{F(r,\theta )}{f(r)}},
\end{equation}
\begin{equation}\label{IC3}
	p_\theta=r \sqrt{F(r,\theta )}\sin\alpha,
\end{equation}
\begin{equation}\label{IC4}
	L=\frac{r \cos\alpha \sin \beta \sin \theta}{\sqrt{F(r,\theta )}},
\end{equation}
where  $\alpha$, $\beta$ are the observation angles and Eqs.~\eqref{IC1}-\eqref{IC4} are evaluated at the observer coordinates. These initial conditions are obtained by projecting the photon's momentum into the observer's tetrad frame - see Ref.~\cite{da2015black} for more details.

The initial conditions of the photon are defined by the observation angles. Using the PyHole package implemented in Python~\cite{Cunha:2016bjh}, varying the observation angles, we evolved 1024$\times$1024 light ray trajectories. An illustration of the backward ray tracing method is depicted in Fig.~\ref{BRT}. Each null geodesic motion contributes to the lensed image with one pixel. If the geodesic is captured by the event horizon, the pixel is assigned with the color black, otherwise we assign a colored pixel. To gain a deeper understanding of the gravitational lensing effect introduced by the presence of the SBHSU, we partitioned the celestial sphere into four quadrants, each designated by a distinct color: red, green, blue, and yellow. In addition to dividing the celestial sphere into four colors, we also insert, on the celestial sphere, a white filled circle right in front of the observer, which will be behind the BH. This circle is usually associated with the appearance of the so called Einstein ring in the lensed image, and gives a better understanding of the gravitational lensing effects.

\subsection{Equatorial plane images}
The shadows and gravitational lensing of the SBHSU, described by Eq.~\eqref{BHmetric}, are displayed in Fig.~\ref{Shadow}. 
We considered the observer positioned on the equatorial plane, $\theta=\pi/2$, and with radial coordinate $r=15M$. We computed the images for several values of the swirling background parameter, namely, $j M^2= n\delta j$, with $n\in\{0,1,2,3,4,5\}$ and $\delta j=0.0001$. The top left image corresponds to the Schwarzschild BH, for reference. We increase the parameter $j$ from left to right, top to bottom, in each image of Fig.~\ref{Shadow}. The contour of the shadow gets a tilted oblate shape in comparison to the Schwarzschild BH. Regarding the gravitational lensing effects in the images, they also become more oblate as $j$ increases. The Einstein ring, present in the Schwarzschild's case, disappears when we increase the swirling parameter.

\subsection{Off-equatorial plane images}
Next we analyze the shadow and gravitational lensing when the observer is positioned off the equator. For this case, we considered images with $jM^2=0.0005$, keeping the same radial distance, but now changing observer $\theta$ coordinate, starting at $\theta=0$ and ending at $\theta=\pi$. We considered equal step displacements $\delta \theta=\pi/8$ in the observer's $\theta$ coordinate.

The simulated images are displayed in Fig.~\ref{Shadow2}. The first (last) image corresponds to an observer located right above (below) the BH. In this case, due to axial symmetry, the shadow's boundary is a perfect circle. The predominant colors for this position of the observer are blue and yellow (green and red), which compose the bottom (top) hemisphere of the celestial sphere, hence, are in front of the observer. Nevertheless, it is still possible to see the other two colors near the shadow's edge, which are on the back of the observer.

As we move the observer along the $\theta$ direction, from the poles to the equator, all the colors blend together and the distinct circular shape of the shadow's edge fades away. Interestingly, moving the observer away from the poles disrupts the $\mathbb{Z}_2$ symmetry. 
The $\mathbb{Z}_2$ symmetry breaking of the shadow is depicted in Fig.~\ref{Shadow3}. 
To represent the reflections we considered 3 different points, namely $P_1$, $Q_1$ and $R_1$; on the shadow's edge in each panel. 
The points $P_1$, $Q_1$ and $R_1$ are 
first reflected with respect to the horizontal axis and end up on the points $P_2$, $Q_2$ and $R_2$, respectively. Afterwards, we consider a second reflection, now with respect to the vertical axis, mapping $P_2$, $Q_2$ and $R_2$ to $P_3$, $Q_3$ and $R_3$, respectively. 
The point $P_1$ is always chosen to be near the vertical axis, the point $R_1$ is close to horizontal axis, whereas the point $Q_1$ is chosen about half the way between the other two.

For an observer with $\theta=0$ or $\theta=\pi$, the shadow's edge has a $\mathbb{Z}_2\times\mathbb{Z}_2$ symmetry. This means that any point on the shadow's boundary, if reflected with respect to either horizontal or vertical axis, ends up in an another point contained on the shadow's edge. Those reflections are illustrated on the top left panel of Fig.~\ref{Shadow3}.

Considering now an observer at $\theta=\pi/4$, no reflection symmetry with respect to the vertical and/or horizontal axes is present. The top right panel in Fig.~\ref{Shadow3} represents the reflection for this case. The point $P_1$ close to vertical axis is (approximately) mapped to another point on the shadow's edge after both vertical and horizontal reflections. However, as we consider points on the shadow closer to the horizontal axis, such as $Q_1$ and $R_1$, the reflections do not map those points to another point on the shadow's boundary. The point that most clearly illustrates the symmetry breaking is $R_1$, located near the horizontal axis.

Finally we considered the case where the observer is at the equatorial plane $\theta=\pi/2$. In this case the odd $\mathbb{Z}_2$ (see Appendix~\ref{Appendix}) is present. Therefore, any point on the shadow's edge, after both vertical and horizontal reflections are mapped to other points on the shadow's edge. Hence, for those observers, the symmetry of the underlying spacetime is inherited for its shadow. Such modification of the symmetry properties of the shadow illustrated here emphasizes the already known concept that the shadow is observer dependent. 

\section{Final Remarks}
\label{FR}
The SBHSU~\cite{Gibbons:2013yq,Astorino:2022aam} represents a novel class of exact solutions of vacuum Einstein's GR, initially explored in Ref.~\cite{Astorino:2022aam} and exhibiting intriguing properties. 
We explored the shadows and gravitational lensing of the SBHSU. Because the SBHSU described with $(t,r,\theta,\varphi)$ coordinates does not appear to be integrable, we employed the backward ray tracing technique to compute the shadow and gravitational lensing. Our findings show that when observers are positioned at the equatorial plane, the shadow inherits the odd reflection symmetry present in the spacetime, giving a prolate shape for its boundary along a diagonal direction - a sort of twisting of the shadow. However, for observers off the equatorial plane, this odd symmetry is lost, except for those situated at $\theta=0$ or $\theta=\pi$, where the shadow exhibits $\mathbb{Z}_2\times\mathbb{Z}_2$ symmetry.

Our analysis of the null geodesic flow in the SBHSU also revealed the existence of two counter-rotating unstable LRs (total topological charge $w=-1$ for each $H_\pm$), each co-rotating with the underlying spacetime. We have shown that, due to the odd $\mathbb{Z}_2$ symmetry, these LRs must exist outside the equatorial plane.

One may compare the number of LRs of the SBHSU with that for the Schwarzschild Melvin solution. When we nonlinearly bring together the Schwarzschild BH ($w=-1$) and the SU ($w=0$), the result is a spacetime with $w=-1$. On the other hand, both Schwarzschild and Melvin spacetimes admit the existence of LRs\footnote{For the Melvin universe, due to the translational invariance along the $z$ direction, there exists a light tube instead of a LR.}. When these two spaces are nonlinearly superposed in the Schwarzschild-Melvin solution, if the magnetic field is weak, there are two LRs on the spacetime, which were inhered by each one of the separate spacetimes. If we let the magnetic field become strong, the potential's critical points are spoiled~\cite{Junior:2021dyw}. By these two examples, heuristically, it seems that the number of LRs, as well as their respective stabilities, of each spacetime is preserved even after the nonlinear superposition, as long as the boundary conditions are not significantly altered. Another example where this summation rule can fail due to boundary conditions is in the Schwarzschild-dilatonic-Melvin BH~\cite{Junior:2021svb}. But, of course, it is hard to formalize this argument, since GR's field equations are nonlinear and, therefore, the associated space of solutions does not constitute a vector space.

The concepts of odd and even $\mathbb{Z}_2$ symmetry, introduced in Appendix~\ref{Appendix}, are fundamental to our results. This classification is important to understand the behavior of null geodesics in the Schwarzschild BH within the SU. Several distinguishing properties of LRs, shadows and gravitational lensing of the SBHSU are related to the odd type of symmetry.

We remark that when delving into the study of exact solutions, one typically encounters scenarios where the spacetime is either overly idealized, making it inadequate for describing real physical systems, or it exhibits pathologies that render its existence in the natural world highly unlikely. During the GR7 international conference in 1974, Kinnersley highlighted this issue: ``the study of exact solutions has acquired a rather low reputation in the past, for which there are several explanations. Most of known exact solutions describe situations which are frankly unphysical and these do have a tendency to distract attention from useful ones''~\cite{Kinnersley:1974xut}. 

The SBHSU spacetime, even though it is free of conical singularities and closed timelike curves, does not have much astrophysical appeal, since it is not asymptotically flat. However, there is also a case to be made if the SBHSU can represent, in some regime of approximation, an analytical model for some astrophysical phenomenon. The authors in Ref.~\cite{Astorino:2022aam} conjectured that the SBHSU might be useful to describe the collision of oppositely rotating galaxies. This issue holds the potential for a standalone work if thoroughly explored.

Apart from the phenomenology perspective,  there exists a significant value in exploring exact solutions for understanding GR in its fully nonlinear regime. Delving into the properties of a specific analytical solution can illuminate entirely novel facets of BHs, exemplified by the shadow profile we have unveiling in this study for the SBHSU. Exact solutions are also important to establish counter examples to seemly truth statements, that are generically false, like the existence of a well-defined equator which does not constitutes a totally geodesics submanifold (see Appendix~\ref{Appendix}). Moreover, there remains a wealth of uncharted territory awaiting exploration within the SBHSU metric (and others), extending far beyond the intricacies of its null geodesic structure.

\appendix
\section{General considerations regarding $\mathbb{Z}_2$ symmetry and null geodesic motion}\label{Appendix}
\subsection{Definition of odd $\mathbb{Z}_2$ symmetry}
The goal of this Appendix is to define the odd $\mathbb{Z}_2$ symmetry and put it in contrast with the usual (even) $\mathbb{Z}_2$ symmetry. The $\mathbb{Z}_2$ symmetry is associated with a particular type of symmetry operation involving some kind of inversion of a physical system. This type of transformation is mathematically represented by the action of the cyclic group of order 2, $\mathbb{Z}_2$~\cite{johnson2018geometries}. 

Let $(\mathcal{M},g_{\mu\nu})$ be 4-dimensional, stationary, axisymmetric, circular spacetime. For this class of spacetimes, we can parameterize the non-Killing submanifold by the coordinates $r\in(r_h,\infty)$ and $\theta\in(0,\pi)$, where $r=r_h$ is the horizon surface. Under these assumptions, the spacetime metric can be written as
\begin{equation}\label{ds}
	ds^2=g_{tt}dt^2+2g_{t\varphi}dtd\varphi+g_{\varphi\varphi}d\varphi^2+g_{rr}dr^2+g_{\theta\theta}d\theta^2.
\end{equation}

In the context of BH physics, or more generically, of localized compact objects, the action of $\mathbb{Z}_2$ is commonly related with the reflection transformation $\theta\rightarrow\theta_R:=\pi-\theta$, which intuitively amounts to changing the "north" and the "south" hemispheres. Spacetimes that are symmetric under such reflection, i.e.
\begin{equation}\label{even}
	\forall (\mu,\nu): g_{\mu\nu}(r,\theta)=g_{\mu\nu}(r,\theta_R),
\end{equation}
 are called $\mathbb{Z}_2$ symmetric and the fixed point $\theta=\pi/2$ is the equatorial plane. The Kerr spacetime, as well as all spherically symmetric spacetimes are examples of $\mathbb{Z}_2$ symmetric spacetimes. But there are well known non-$\mathbb{Z}_2$ symmetric spacetimes, which have been studied in the context of light propagation~\cite{Larsen:1999pp,Cunha:2019dwb,Chen:2020aix}.

The SBHSU is not $\mathbb{Z}_2$ symmetric, but it has the discrete symmetry defined by 
\begin{equation}\label{symmetry}
	(t,\theta)\rightarrow(-t,\theta_R).
\end{equation}
That is, the metric is invariant under exchanging north and south \textit{and} simultaneously the direction of the time coordinate. 
We will dub the spacetime with the symmetry defined by Eq.~\eqref{symmetry} as \textit{odd} $\mathbb{Z}_2$ \textit{symmetric}, whereas the standard $\mathbb{Z}_2$ symmetry will be referred as \textit{even} $\mathbb{Z}_2$ \textit{symmetry}. The SBHSU is also invariant under the same discrete transformation as (say) Kerr, namely $(t,\varphi)\rightarrow(-t,-\varphi)$; its distinctive feature is that it is \textit{not} invariant under the standard (even) $\mathbb{Z}_2$ symmetry $\theta\rightarrow \theta_R$; it requires the additional transformation present in Eq.~\eqref{symmetry}.

For odd symmetric spacetimes, we can still define the equatorial plane as the set of points which are fixed by the $\theta$-reflection. Hence, the top submanifold $0<\theta<\pi/2$ is isometric do the bottom spacetime patch $\pi/2<\theta<\pi$, under time reversal for one of them. Equation~\eqref{symmetry} could also be written with a reflection in $\varphi$ instead of $t$. The symmetry defined in Eq.~\eqref{symmetry} holds true for any spacetime with metric components invariant under $\theta$-reflection, except for the $g_{t\varphi}$ component, which must pick up a minus sign. Therefore, an equivalent definition of an odd $\mathbb{Z}_2$ (symmetric) spacetime can be given, in the above metric chart, in terms of metric components transformations as follows:
\begin{equation}\label{symmetry1}
	\forall (\mu,\nu)\neq (t,\varphi): g_{\mu\nu}(r,\theta)=g_{\mu\nu}(r,\theta_R),
\end{equation}
\begin{equation}\label{symmetry2}
	 g_{t\varphi}(r,\theta)=-g_{t\varphi}(r,\theta_R).
\end{equation}

Thus, the reflection in $t$ (or $\varphi$) corrects the ``wrong'' sign of the $g_{t\varphi}$ component. Moreover, any spacetime with the symmetry defined by Eq.~\eqref{symmetry} must have $g_{t\varphi}|_{\theta=\pi/2}=0$, since $g_{t\varphi}$ is an odd function with respect to the plane $\theta=\pi/2$.

Apart from the SBHSU, another example of spacetime that is odd $\mathbb{Z}_2$ symmetric is the Taub-NUT BH.

\subsection{The potentials $H_{\pm}$}
It is possible to express the potentials $H_{\pm}$ purely in terms of the metric components. Thus, we have
\begin{equation}\label{HpmG}
	H_{\pm}=\frac{-g_{t\varphi}\pm\sqrt{g_{t\varphi}^2-g_{tt}g_{\varphi\varphi}}}{g_{\varphi\varphi}}.
\end{equation} 
One may check that Eq.~\eqref{HpmG} agrees with Eq.~\eqref{Hpm} when the metric components are given by Eq.~\eqref{BHmetric}.

The symmetry relation in Eq.~\eqref{Hsym} between the potentials $H_{\pm}$ is true for any spacetime that is symmetric under action of Eq.~\eqref{symmetry}. This can be shown as follows:
\begin{equation}\label{Z2prop}
	\begin{aligned}
		&H_{\pm}(r,\theta)=\frac{-g_{t\varphi}(r,\theta)\pm\sqrt{g_{t\varphi}(r,\theta)^2-g_{tt}(r,\theta)g_{\varphi\varphi}(r,\theta)}}{g_{\varphi\varphi}(r,\theta)}\\
		&=-\frac{-g_{t\varphi}(r,\theta_R)\mp\sqrt{g_{t\varphi}(r,\theta_R)^2-g_{tt}(r,\theta_R)g_{\varphi\varphi}(r,\theta_R)}}{g_{\varphi\varphi}(r,\theta_R)}\\
		&=-H_{\mp}(r,\theta_R).
	\end{aligned}
\end{equation}

Eq.~\eqref{Z2prop} shows that, for odd $\mathbb{Z}_2$ symmetric spacetimes, the potential $H_{-}$ can be fully constructed from the $H_+$ potential and vice-versa, which is not true for even $\mathbb{Z}_2$ symmetric spacetimes. 

\subsection{Equatorial light rings}
It is interesting to investigate the relation between even $\mathbb{Z}_2$ symmetry and LRs positioned on the equator, assuming a single BH with spherical topology. This can be simply addressed for spacetimes possessing the same properties mentioned in the preceding section plus asymptotically flatness. 

Let $\tilde{H}_{\pm}$ be null geodesic potentials of a BH with such properties. LRs are critical points of the potentials $\tilde{H}_{\pm}$, i.e. the LR position $(\tilde{r}_\pm,\tilde{\theta}_\pm)$ is defined by $\nabla \tilde{H}_\pm(\tilde{r}_\pm,\tilde{\theta}_\pm)=0$. From the first three hypotheses we obtain
\begin{equation}\label{LRcond}
	\forall\theta\in \mathscr{I} \subset(0,\pi)\ \exists\ \tilde{r}_{\pm}\in(r_h,\infty):\ \frac{\partial \tilde{H}_{\pm}(\tilde{r}_{\pm},\theta)}{\partial r} =0,
\end{equation}
where $\mathscr{I}$ is an open interval that includes $\pi/2$. In fact, for asymptotically flat spacetimes, a stronger version of the condition~\eqref{LRcond} is satisfied, which is valid for all values of $\theta\in(0,\pi)$ and not only for an open neighborhood of $\pi/2$. With this assumption, one can also show that there is a odd number of $\tilde{r}_{\pm}$'s satisfying Eq.~\eqref{LRcond}.

Now, imposing the even $\mathbb{Z}_2$ symmetry, we obtain that
\begin{equation}\label{Z2cond}
	\forall\theta\in \mathscr{I} \subset(0,\pi):\ \frac{\partial \tilde{H}_{\pm}(r,\theta)}{\partial \theta} =-\frac{\partial \tilde{H}_{\pm}(r,\pi-\theta)}{\partial \theta}.
\end{equation}

In particular, we may choose $\theta=\pi/2$ in Eq.~\eqref{Z2cond}, from where we get $\partial \tilde{H}_{\pm}(r,\pi/2)/\partial \theta =0$. Therefore, the following ``formal'' implication is true:
\begin{equation}\label{implication1}
	\text{even }\mathbb{Z}_2\Rightarrow \exists \text{ LR at }\theta=\pi/2.
\end{equation}

Alternatively, we could have calculated the gradient of Eq.~\eqref{HpmG} and evaluated at the equatorial plane. If the spacetime is even $\mathbb{Z}_2$ symmetric, the metric components should be even functions with respect to the equator, according to Eq.~\eqref{even}. Thus, their corresponding derivatives with respect to $\theta$ are odd functions, which implies Eq.~\eqref{implication1}.

It also holds the contrapositive of Eq.~\eqref{implication1}, which states that the absence of a LR at $\theta=\pi/2$ implies that the corresponding spacetime is not even $\mathbb{Z}_2$ symmetric, i.e.
\begin{equation}\label{implication2}
	(\text{$\nexists$ LR at }\theta=\pi/2) \Rightarrow 	\neg(\text{even }\mathbb{Z}_2).
\end{equation}

We remark, however, that  \textit{it is not true} in general that: ($\exists$ LR at $\theta=\pi/2$) $\Rightarrow \text{even }\mathbb{Z}_2$, since is easy to construct examples of spacetimes with a LR at the equatorial plane, which are not even $\mathbb{Z}_2$ symmetric. Both Eqs.~\eqref{implication1} and~\eqref{implication2} are only valid for asymptotically flat spacetimes.

Therefore, at least for the asymptotically flat cases, the relation between even $\mathbb{Z}_2$ symmetry and equatorial LRs - assuming a single BH\footnote{The point here is that for, say, a 2-centre solution, $e.g.$ the 2-centre Majumdar-Papapetrou solution, there needs not to be a LR on the equatorial plane. The loophole in such cases is because the horizon is multi-connected, and therefore it does not have spherical topology as assumed in Ref.~\cite{Cunha:2020azh} to compute the topological charge; it is rather a product of spheres.} - is already well established. For other asymptotics,  Eq.~\eqref{LRcond} is not necessarily true. For instance, considering a BH which is asymptotically Melvin, the existence of critical points in the radial direction is determined by the strength of the magnetic field. If the magnetic field is weak (subcritical regime), then Eq.~\eqref{LRcond} is satisfied, but for strong magnetic fields (supercritical regime) it is not~\cite{Junior:2021dyw}. Nevertheless, the condition~\eqref{LRcond} is necessary in order to have an equatorial LR in the first place, thus the general conclusion is: \textit{for any stationary, axisymmetric, even $\mathbb{Z}_2$ symmetric, (single) BH spacetime where Eq.~\eqref{LRcond} is satisfied, there exists one LR lying on the equatorial plane.}

As we shall see, when we assume odd $\mathbb{Z}_2$ symmetry, the result is different. The odd symmetry implies that the metric components should transform according to Eqs.~\eqref{symmetry1} and~\eqref{symmetry2}. Thus, we have that
\begin{equation}\label{symmetry3}
	\forall (\mu,\nu)\neq (t,\varphi): \frac{\partial g_{\mu\nu}(r,\theta)}{\partial \theta}=-\frac{\partial g_{\mu\nu}(r,\theta_R)}{\partial \theta},
\end{equation}
\begin{equation}\label{symmetry4}
	 \frac{\partial g_{t\varphi}(r,\theta)}{\partial \theta}=\frac{\partial g_{t\varphi}(r,\theta_R)}{\partial \theta}.
\end{equation}
From Eqs.~\eqref{symmetry3} and~\eqref{symmetry4} we may infer that $\forall (\mu,\nu)\neq (t,\varphi):\partial g_{\mu\nu}(r,\pi/2)/\partial \theta=0$, but nothing can be concluded for the component $g_{t\varphi}(r,\pi/2)$.

Let $\bar{H}_{\pm}$ be null geodesic potentials of an odd $\mathbb{Z}_2$ symmetric BH. Using Eqs.~\eqref{symmetry2} and~\eqref{symmetry3}, we have that
\begin{equation}
	\frac{\partial \bar{H}_{\pm} (r,\pi/2)}{\partial \theta}=-\frac{\partial g_{t\varphi}(r,\pi/2)/\partial \theta}{g_{\varphi\varphi}\left(r,\pi/2\right)} .
\end{equation}

 As we did for the even-symmetric case, we may assume that Eq.~\eqref{LRcond} is true for $\bar{H}_{\pm}$. Hence, we have a condition for the existence of LRs at the equator for odd $\mathbb{Z}_2$ spacetimes, which is given by
 \begin{equation}\label{odd}
 	\frac{\partial g_{t\varphi} (r,\pi/2)}{\partial \theta}=0.
 \end{equation}

The interpretation of Eq.~\eqref{odd} is that the spacetime, despite the fact that it is odd $\mathbb{Z}_2$ symmetric, is locally even $\mathbb{Z}_2$ symmetric in a small neighborhood of the equator. For this to be accomplished, the $g_{t\varphi}$, which vanishes at the equatorial plane (see Eq.~\eqref{symmetry2}), is also zero in a small vicinity of $\theta=\pi/2$. As we shall see in the next subsection, Eq.~\eqref{odd}) is also a condition for the equatorial plane to be a totally geodesic submanifold of an odd $\mathbb{Z}_2$ symmetric spacetime. Therefore, the existence of equatorial LRs for odd symmetric BHs is intimately related with how null geodesics deviate from the equatorial plane.

From Eq.~\eqref{BHmetric}, one can show that
\begin{equation}
	\frac{\partial g_{t\varphi} (r,\pi/2)}{\partial \theta}=-\frac{4 j r^2 (r-2 M)}{j^2 r^4+1}\neq 0,
\end{equation}
which shows that the SBHSU spacetime cannot have LRs at the equatorial plane. A similar analysis can be done for the Taub-NUT BH.

\subsection{Equatorial totally geodesic submanifold}

\begin{figure}
	\centering
	\includegraphics[width=\columnwidth]{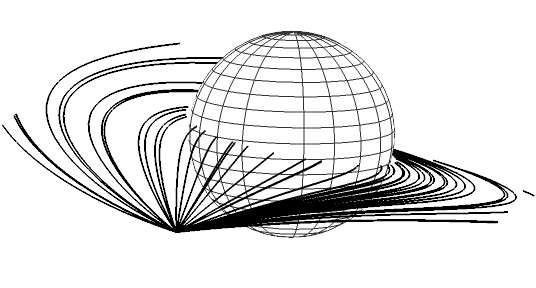}
	\includegraphics[width=\columnwidth]{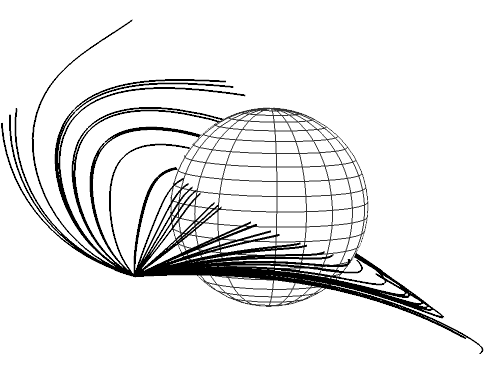}
	\includegraphics[width=\columnwidth]{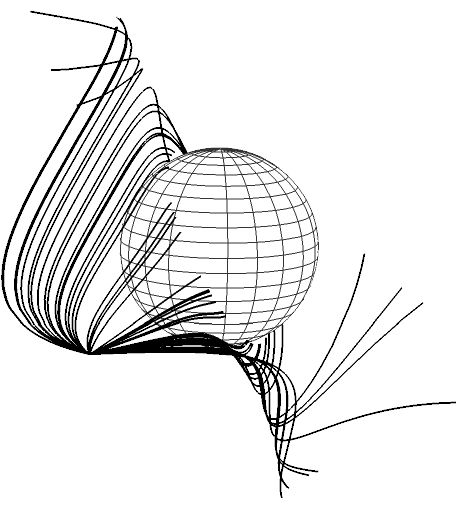}
	\caption{Deviation of geodesics from the equatorial plane. We evolved, numerically, 50 null geodesics with initial conditions given by Eqs.~\eqref{IC1}-\eqref{IC4},  setting $\alpha=0$, while $\beta$ varies randomly.}
	\label{Ap}
\end{figure}

Let $\bar{\mathcal{M}}$ denote a submanifold of $\mathcal{M}$, defined by $\theta=\pi/2$. 
The induced metric $h_{\mu\nu}$ on $\bar{\mathcal{M}}$ (also known as the first fundamental form) is given by the expression $h_{\mu\nu}=g_{\mu\nu}-n_\mu n_\nu$. Let $n=\left(1/\sqrt{g_{\theta\theta}}\right)\partial_\theta$ be the normal unit vector to $\bar{\mathcal{M}}$. The extrinsic curvature $k_{\mu\nu}$ (or the second fundamental form) of the hypersurface $\bar{\mathcal{M}}$ is the symmetric tensor defined by
\begin{equation}
	k_{\mu\nu}=\frac{1}{2}\mathscr{L}_n h_{\mu\nu},
\end{equation}
where $\mathscr{L}_n$ denotes the Lie derivative with respect to the normal vector $n$~\cite{Carroll:2004st}. The spacetime patch $\bar{\mathcal{M}}$ of $\mathcal{M}$ is said to be a totally geodesic submanifold if observers in $\mathcal{M}$ see no curving in $\bar{\mathcal{M}}$, i.e. $k_{\mu\nu}=0$~\cite{o1983semi}.

For a spacetime defined by the metric~\eqref{ds} we can calculate the extrinsic curvature explicitly. The pullback of the extrinsic curvature $k_{ab}=\left(\partial x^\mu/\partial x^a\right)\left(\partial x^\nu/\partial x^b\right)k_{\mu\nu}$\footnote{Here, we are working under the assumption that Latin indices are limited to the set $\{t, r, \varphi\}$.} is given by
\begin{equation}\label{kab}
	\left(k_{ab}\right)=\frac{1}{2\sqrt{g_{\theta\theta}}}\begin{pmatrix}
		\partial_\theta g_{tt} & 0 & \partial_\theta g_{t\varphi} \\
		0 & \partial_\theta g_{rr} & 0 \\
		\partial_\theta g_{t\varphi} & 0 & \partial_\theta g_{\varphi\varphi}
	\end{pmatrix}.
\end{equation}

For even $\mathbb{Z}_2$ symmetric spacetimes we have that
\begin{equation}\label{evenR}
	\begin{aligned}
		g_{\mu\nu}(\theta)=g_{\mu\nu}(\theta_R)&\Rightarrow\partial_\theta g_{\mu\nu}(\theta)=-\partial_\theta g_{\mu\nu}(\theta_R)\\
		&\Rightarrow\partial_\theta g_{\mu\nu}(\pi/2)=0,
	\end{aligned}
\end{equation}
 hence Eq.~\eqref{kab} vanishes, which satisfy the condition of extrinsic flatness.

Similarly, for odd $\mathbb{Z}_2$, one can infer that 
\begin{equation}
	\begin{aligned}
		&\underset{(\mu,\nu)\neq(t,\varphi)}{g_{\mu\nu}(\theta)}=\underset{(\mu,\nu)\neq(t,\varphi)}{g_{\mu\nu}(\theta_R)}\Rightarrow \underset{(\mu,\nu)\neq(t,\varphi)}{\partial_\theta g_{\mu\nu}(\pi/2)}=0,\\
		&g_{t\varphi}(\theta)=-g_{t\varphi}(\theta_R)\Rightarrow \partial_\theta g_{t\varphi}(\theta)=\partial_\theta g_{t\varphi}(\theta_R).
	\end{aligned}
\end{equation}

Therefore, the difference in this case is that the $\partial_\theta g_{t\varphi}(\theta)$ term does not necessarily vanish at the equatorial plane, which implies that, generically, for odd symmetric spacetimes, $k_{ab}\neq0$. Nevertheless, those spacetimes can have an equatorial totally geodesic submanifold, as long as Eq.~\eqref{odd} is satisfied. In general, a necessary and sufficient condition for a spacetime (with metric given by Eq.~\eqref{ds}) to have an equatorial totally geodesic submanifold is to be locally even $\mathbb{Z}_2$ around the plane $\theta=\pi/2$. 

An equivalent definition of totally geodesic submanifold is given in terms of geodesics. Let $T_p\mathcal{M}$ denote the tangent space of $\mathcal{M}$ at the point $p\in\mathcal{M}$. If, in a small interval $\lambda\in(-\epsilon,\epsilon)\subset\mathbb{R}$, a geodesic $\gamma(\lambda)$ of $\mathcal{M}$ with tangent vector $v\in T_{\gamma(\lambda)}\mathcal{M}$ have initial conditions given by $\gamma(0)\in\bar{\mathcal{M}}$ and $v\in T_{\gamma(0)}\bar{\mathcal{M}}$, lies in $\bar{\mathcal{M}}$, then $\bar{\mathcal{M}}$ is classified as a totally geodesic submanifold. Thus, every geodesic of $\bar{\mathcal{M}}$ must also be a geodesic of $\mathcal{M}$~\cite{o1983semi}.

To ensure consistency, let us demonstrate that a brief exploration of the geodesic equation results in identical criteria for the presence of an equatorial totally geodesic submanifold. Considering a spacetime characterized by the line element described in Eq.~\eqref{ds}, we can derive the geodesic equation specifically for the $\theta$ coordinate as follows:
\begin{equation}\label{GeodEq}
	\ddot{\theta}+\Gamma^\theta_{tt}\dot{t}^2+\Gamma^\theta_{rr}\dot{r}^2+\Gamma^\theta_{\theta\theta}\dot{\theta}^2+\Gamma^\theta_{\varphi\varphi}\dot{\varphi}^2+2\Gamma^\theta_{t\varphi}\dot{t}\dot{\varphi}+2\Gamma^\theta_{r\theta}\dot{r}\dot{\theta}=0,
\end{equation}
where
\begin{equation}\label{Gamma}
	\begin{aligned}
		\Gamma^\theta_{tt}=-\frac{\partial_\theta g_{tt}}{2 g_{\theta\theta}},\ \ \ \  & \Gamma^\theta_{\varphi\varphi}=-\frac{\partial_\theta g_{\varphi\varphi}}{2 g_{\theta \theta}} ,\ \ \ \ \Gamma^\theta_{t\varphi}=-\frac{\partial_\theta g_{t\varphi}}{2 g_{\theta \theta}} , \\
		\Gamma^\theta_{rr}=-\frac{\partial_\theta g_{rr}}{2 g_{\theta\theta}},\ &\ \ \   \Gamma^\theta_{\theta\theta}=\frac{\partial_\theta g_{\theta\theta}}{2 g_{\theta \theta}} ,\ \ \ \ \Gamma^\theta_{r \theta}=\frac{\partial_r g_{\theta \theta}}{2 g_{\theta \theta}}.
	\end{aligned}
\end{equation}

In order to set the particle movement at the equator, we choose the initial condition 
\begin{equation}
	\theta=\pi/2,\ \ \ \ \dot{\theta}=0,
\end{equation}
which eliminates the terms $\Gamma^\theta_{\theta\theta}$ and  $\Gamma^\theta_{r\theta}$. The remaining connection terms are precisely given by the components of the extrinsic curvature tensor $k_{ab}$. Therefore, the analysis is identical as the one made before, as expected from the equivalence of the definitions. We could also have written the extrinsic curvature with the equivalent formula~\cite{rahaman2021general}
\begin{equation}\label{kab2}
	\begin{aligned}
		k_{ab}&=-n_\mu\left(\frac{\partial^2 x^\mu}{\partial x^a \partial x^b}+\Gamma^{\mu}_{\alpha \beta}\frac{\partial x^\alpha}{\partial x^a}\frac{\partial x^\beta}{\partial x^b}\right)=-\sqrt{g_{\theta\theta}}\Gamma^\theta_{ab},
	\end{aligned}
\end{equation}
which is in accordance with Eqs.~\eqref{kab} and~\eqref{Gamma}. 

We conclude that, for odd $\mathbb{Z}_2$ symmetric spacetimes, it is possible for geodesics initially confined to the plane $\theta=\pi/2$, to deviate from it, namely
\begin{equation}
	g_{\theta\theta}\ddot{\theta}=\dot{t}\dot{\varphi}\partial_\theta g_{t\varphi}.
\end{equation}

In the literature, the existence of a well defined equator is associated only with spacetimes which are even $\mathbb{Z}_2$ symmetric. Expanding on this notion and from the results obtained in Eq.~\eqref{evenR}, one might think that the existence of a well defined equator is linked to a totally geodesic submanifold, which is not true. Our demonstration revealed that odd $\mathbb{Z}_2$ symmetric spacetimes provide a counter example, since they can also have a well defined equatorial plane, while also having geodesics escaping from it.

To illustrate how the geodesics deviate from the equatorial plane, we considered a plot of 50 geodesics in the SBHSU spacetime, with initial conditions given by Eqs.~\eqref{IC1}-\eqref{IC4}, $\theta=\pi/2$ and $r=10M$. We fix $\alpha=0$, which corresponds to geodesics launched at the equatorial plane ($p_\theta=0$), and let $\beta$ take values randomly. The plots are displayed in Fig.~\ref{Ap} for $jM^2=0.001,\ 0.0025,\ 0.01$.

For the SBHSU, there exists a critical value for $j$, namely $j_\text{c}=r^{-2}$, with $r$ being the observer radial position, such that $|\partial_\theta g_{t\varphi}|$ is maximum. Therefore, the deviation of the geodesics is amplified for $j=j_\text{c}$. For $r=10M$, we have $j_\text{c}M^2=0.01$, which corresponds to the bottom plot in Fig.~\ref{Ap}. For $j>j_\text{c}$, the geodesic deviation from the equator starts to diminish. For $j\to\infty$, we have $\partial_\theta g_{t\varphi}\to0$, so that, for high enough values of $j$, this effect disappears. 

\begin{acknowledgments}
	We would like to thank Pedro Cunha and Marco Astorino for discussions and comments on a draft of this work. We also thank Rogério Capobianco and Betti Hartmann for pointing out inconsistencies in Figs.~\ref{Shadow} and~\ref{Shadow2} in the first version of this paper. They have been corrected, without changing any conclusions. We are grateful to Funda\c{c}\~ao Amaz\^onia de Amparo a Estudos e Pesquisas (FAPESPA), Conselho Nacional de Desenvolvimento Cient\'ifico e Tecnol\'ogico (CNPq) and Coordena\c{c}\~ao de Aperfei\c{c}oamento de Pessoal de N\'ivel Superior (CAPES) -- Finance Code 001, from Brazil, for partial financial support. 
	ZM and LC thank the University of Aveiro, in Portugal, for the kind hospitality during the completion of this work.
	This work is supported  by the  Center for Research and Development in Mathematics and Applications (CIDMA) through the Portuguese Foundation for Science and Technology (FCT -- Fundaç\~ao para a Ci\^encia e a Tecnologia), references  UIDB/04106/2020 and UIDP/04106/2020.  
	The authors acknowledge support  from the projects CERN/FIS-PAR/0027/2019, PTDC/FIS-AST/3041/2020, CERN/FIS-PAR/0024/2021 and 2022.04560.PTDC.  
	This work has further been supported by  the  European  Union's  Horizon  2020  research  and  innovation (RISE) programme H2020-MSCA-RISE-2017 Grant No.~FunFiCO-777740 and by the European Horizon Europe staff exchange (SE) programme HORIZON-MSCA-2021-SE-01 Grant No.~NewFunFiCO-101086251.
\end{acknowledgments}

\bibliography{ref}

\end{document}